\newif\iffullversion{}
\fullversiontrue{}

\iffullversion%
\documentclass[a4paper]{article}
\else%
\documentclass[USenglish,oneside,twocolumn]{article}
\fi

\iffullversion%
\usepackage{amsmath}
\usepackage[margin=1in]{geometry}
\else%
\usepackage[utf8]{inputenc}
\usepackage[big]{dgruyter_NEW}
\cclogo{\includegraphics{by-nc-nd.pdf}}
\fi%

\usepackage{mathtools}
\usepackage{amssymb}
\usepackage{amsthm}
\usepackage{amsfonts}
\usepackage{graphicx}
\usepackage[pdf]{graphviz}
\usepackage{enumitem}
\usepackage{multirow}
\usepackage{dblfloatfix}
\usepackage{float}
\usepackage{booktabs}
\usepackage{xspace}
\usepackage{subfig}
\usepackage[group-separator={,}]{siunitx}
\PassOptionsToPackage{hyphens}{url}\usepackage{hyperref}

\usepackage[pdftex,dvipsnames]{xcolor}
\usepackage[colorinlistoftodos]{todonotes}

\usepackage{verbatim}
\usepackage{mathtools}

\usepackage{pgfplots}

\setlist[itemize]{leftmargin=0.5cm}
\setlist[enumerate]{leftmargin=0.5cm}
\theoremstyle{plain}

\newtheorem{theorem}{Theorem}
\newtheorem*{theorem*}{Theorem}
\newtheorem{lemma}{Lemma}
\newtheorem*{lemma*}{Lemma}
\newtheorem{corollary}{Corollary}
\newtheorem*{remark}{Remark}
\newtheorem{example}{Example}
\newtheorem{definition}{Definition}
\newtheorem*{definition*}{Definition}
\newtheorem{prop}{Proposition}
\newtheorem*{mechanism*}{Mechanism}

\iffullversion%
\newcommand{\figurescale}{0.9}
\else
\newcommand{\figurescale}{0.8}
\fi

\newcommand{\numbers}{\mathbb{N}}
\newcommand{\tuples}{\mathcal{T}}
\newcommand{\datasets}{\mathcal{D}}
\newcommand{\knowledges}{\mathcal{B}}
\newcommand{\dk}{\datasets\times\knowledges}

\newcommand{\outputs}{\mathcal{O}}
\newcommand{\mecha}{\mathcal{M}}
\newcommand{\mech}[1]{\mecha\left(#1\right)}
\newcommand{\ind}{\approx}
\newcommand{\indeps}{\ind_{\eps}}
\newcommand{\indepsdel}{\ind_{\eps,\del}}
\newcommand{\PrivLoss}{\mathcal{L}}
\newcommand{\privlossM}[2]{\PrivLoss^{\mecha}_{#1/#2}}
\newcommand{\privloss}[2]{\PrivLoss_{#1/#2}}
\newcommand{\privlossMit}[2]{\PrivLoss^{\mecha,\theta}_{i,#1,#2}}
\newcommand{\privlossit}[3]{\PrivLoss^{#3,\theta}_{i,#1,#2}}

\newcommand{\Simu}{\text{Sim}}
\newcommand{\Sim}[1]{\Simu\left(#1\right)}
\newcommand{\APK}{\text{APK}}
\newcommand{\PPK}{\text{PPK}}
\newcommand{\rrdl}{\mathcal{R}_{2\lambda}}
\newcommand{\ber}[1]{\text{Bernoulli}\left(#1\right)}
\newcommand{\shuffle}{\mathcal{S}}
\newcommand{\shuffl}[1]{\shuffle\left(#1\right)}
\newcommand{\Dep}{\text{Dep}}
\newcommand{\Depab}{\Dep^{\mecha_1,\mecha_2,\theta}_{i,a,b}\left(O_1,O_2\middle|\hatB\right)}

\newcommand{\eps}{\varepsilon}
\newcommand{\del}{\delta}
\newcommand{\epsdel}{{\left(\eps,\del\right)}}

\newcommand{\brackets}[1]{\left[#1\right]}
\newcommand{\bracketscond}[2]{\brackets{#1\middle|#2}}
\newcommand{\Prob}{\mathbb{P}}
\newcommand{\proba}[1]{\Prob\brackets{#1}}
\newcommand{\probac}[2]{\Prob\bracketscond{#1}{#2}}
\newcommand{\probas}[2]{\Prob_{#1}\brackets{#2}}
\newcommand{\probasc}[3]{\Prob_{#1}\bracketscond{#2}{#3}}

\newcommand{\Expect}{\mathbb{E}}

\newcommand{\expectu}[2]{\underset{#1}{\Expect}\brackets{#2}}
\newcommand{\cond}[1]{_{\vert{}#1}}
\newcommand{\Theps}{\left(\Theta,\eps\right)}
\newcommand{\Theeps}{\left(\Theta,2\eps\right)}
\newcommand{\theps}{\left(\left\{\theta\right\},\eps\right)}
\newcommand{\Thepsdel}{\left(\Theta,\eps,\del\right)}
\newcommand{\thepsdel}{\left(\theta,\eps,\del\right)}
\newcommand{\tit}{\theta\in\Theta}
\newcommand{\Darrow}{D_{i\rightarrow{}b}}

\newcommand{\hs}{\hspace{1.5mm}}
\newcommand{\hhs}{\hspace{2cm}}
\newcommand{\mathcmd}[1]{\ensuremath{#1}\xspace}
\newcommand{\supp}{\mathcmd{\mathrm{support}}}
\newcommand{\set}[1]{\mathcmd{\{#1\}}}

\newcommand{\hatD}{\hat{D}}
\newcommand{\hatO}{\hat{O}}
\newcommand{\hatB}{\hat{B}}
\newcommand{\hatheta}{\hat{\theta}}

\newcommand{\hbik}{\hatB\in\knowledges}
\newcommand{\bhb}{B=\hatB}
\newcommand{\dhd}{D=\hatD}

\newcommand{\mohb}{m\left(O,\hatB\right)}
\newcommand{\mob}{m\left(O,B\right)}

\newcommand{\ol}[1]{\overline{#1}}

\begin{document}

\iffullversion%
\title{Differential privacy with partial knowledge}
\fi

\iffullversion%
  \author{%
    \texorpdfstring{%
      \noindent\begin{minipage}{\textwidth}
        \begin{minipage}{0.5\textwidth}
          \centering
          Damien Desfontaines \\
          ETH Zurich / Google \\
          damien@desfontain.es \\
          \leavevmode\\
          Elisabeth Krahmer \\
          LMU Munich \\
          elisabeth.krahmer@campus.lmu.de
        \end{minipage}
        \begin{minipage}{0.5\textwidth}
          \centering
          Esfandiar Mohammadi \\
          University of Lübeck \\
          esfandiar.mohammadi@uni-luebeck.de \\
          \leavevmode\\
          David Basin \\
          ETH Zurich \\
          basin@inf.ethz.ch
        \end{minipage}
      \end{minipage}
    }
    {Damien Desfontaines, Elisabeth Krahmer, Esfandiar Mohammadi, David Basin}
  }
  \maketitle
  \thispagestyle{plain}
  \pagestyle{plain}
\fi


\begin{abstract}
{Differential privacy offers formal quantitative guarantees for algorithms over
  datasets, but it assumes attackers that know and can influence all but one
  record in the database. This assumption often vastly overapproximates the
  attackers' actual strength, resulting in unnecessarily poor utility.
\\
Recent work has made significant steps towards privacy in the presence of
partial background knowledge, which can model a realistic attacker's
uncertainty. Prior work, however,
\iffullversion%
  has definitional problems for correlated data and
\fi%
  does not precisely characterize the underlying attacker model.
\iffullversion%
  We propose a practical criterion to prevent problems due to correlations,
  and we show how to characterize
\fi%
attackers with limited influence or only partial background knowledge over the
dataset. We use these foundations to analyze practical scenarios: we
significantly improve known results about the privacy of counting queries
under partial knowledge, and we show that thresholding can provide formal
guarantees against such weak attackers, even with little entropy in the data.
These results allow us to draw novel links between $k$-anonymity and
differential privacy under partial knowledge. Finally, we prove composition
results on differential privacy with partial knowledge, which quantifies the
privacy leakage of complex mechanisms.
\\
Our work provides a basis for formally quantifying the privacy of many
widely-used mechanisms, e.g.\ publishing the result of surveys, elections or
referendums, and releasing usage statistics of online services.}

\end{abstract}

\iffullversion%
\tableofcontents
\fi%

\section{Introduction}

Differential privacy (DP) is an established privacy notion for algorithms over
datasets. DP models strong attackers that can not only learn, but can even
influence, all but one element from the input dataset. While this strong
attacker model over-approximates realistic attackers, it can also lead to
overly cautious choices of noise parameters, unnecessarily deteriorating the
algorithms' accuracy. Relaxing the assumptions about attackers' background
knowledge and their influence on the data set can lead to smaller noise
parameters and, in turn, to more accurate results.

Consider a national referendum where more than 10 million people vote on a
Yes/No question. Do the exact results of this referendum reveal information
about individuals? It is very reasonable to assume that no realistic attacker
has background knowledge of more than 99\% of all votes. The remaining
uncertainty of 1\% of the data (100k data points) leads to a significant
uncertainty that can, if properly quantified, show that no attacker can use the
results of the referendum to determine how a given individual voted: the
referendum results are private, even if no noise was added to them.

In some scenarios~\cite{venkatadri2018privacy}, one cannot exclude that the
attacker can influence the entire dataset. But there are many natural scenarios
in which the risk of an attacker injecting a large number of data points into
the dataset is negligible: censuses, phone polls, or elections are natural
examples\footnote{Here, we assume the absence of malicious insiders who could
break privacy anyway by leaking the entire dataset.}. In those case, it is
appropriate to consider privacy guarantees that model weaker attackers without
influence over the dataset, but with some background knowledge. While the
precise estimation of an attacker's capabilities may be difficult, a more
balanced privacy analysis should characterize the privacy leakage against
attackers with varying degrees of background knowledge and influence over the
dataset.

As we show in this paper, while there is a rich body of prior
work~\cite{duan2009privacy,rastogi2009relationship,bhaskar2011noiseless,kifer2012rigorous,bassily2013coupled,li2013membership}
on this topic,
\iffullversion%
it fails to account for data with correlations,
\fi%
it does not make the attacker model explicit and precise, and it leaves the
following questions unanswered.
\iffullversion%
What are appropriate notions of privacy for scenarios where the attacker has
partial knowledge about a dataset that contains correlations?
\fi%
How to define notions of privacy under partial knowledge that cleanly delineate
between attackers who only have some background knowledge and attackers who can
influence the data? And can we use these notions in practical contexts, without
making risky assumptions on the data, on how many computations are performed on
the data, or on the adversary's capabilities?

\subsection{Approach and contributions}

In this work, we provide a theoretical foundation for answering these questions,
which we apply to common use cases.
\iffullversion%
Our formalism solves the problems that previous definitions have when the data
contains correlations, and it clearly delineates between attackers that only
have some background knowledge and attackers that can influence the data.
\fi%
We build on these foundations to further analyze common noiseless mechanisms and
prove strong and intuitive results about their privacy. Our main contributions
are as follows.

\iffullversion%
First, we show that existing notions of privacy under partial knowledge break
down when the data contains correlations, allowing very revealing mechanisms to
be mistakenly considered private. We propose a practical criterion and approach
to fix this class of issues.
\fi%

\iffullversion%
Second,
\fi%
we show that there are two distinct ways to model partial knowledge,
depending on whether the attacker can only learn some properties of the data or
can modify the data. We define two corresponding notions of privacy under
partial knowledge: active partial knowledge, where the attacker can influence
the dataset, and passive, where the attacker is unable to influence the dataset.
We show that these notions have natural properties, and prove that they are
equivalent for a large class of common mechanisms and assumptions over the data.
Moreover, we show that the active partial knowledge assumption can be used to
alleviate the challenge of precisely estimating the dataset distribution.

\iffullversion%
Third,
\fi%
putting these results to work, our results provide a formal account of the
privacy of common kinds of queries. We show that counting queries under partial
knowledge can provide privacy, with significantly lower bounds than those
previously given. We also show that thresholding---solely returning the output if
it is larger than a given threshold---can render counting queries private
against passive attackers, even if the input distribution does not have enough enough entropy for
the previous result to apply.

\iffullversion%
Fourth,
\fi%
we prove bounds for the sequential composition of noiseless mechanisms. This
allows us to quantify the privacy leakage of multiple mechanisms with the same
input, or of a mechanism repeated over time.

\subsection{Related work}

Among the numerous variants of differential privacy
(DP)~\cite{desfontaines2019sok}, two main variants model adversaries with
partial background knowledge, using indistinguishability: \emph{noiseless
privacy}~\cite{duan2009privacy,bhaskar2011noiseless}, and \emph{distributional
DP (DDP)}~\cite{bassily2013coupled}.
\iffullversion%
This work discusses the shortcomings of DDP in Section~\ref{sec:correlated}, and
in
\fi%
Section~\ref{sec:passiveactive}, we use the formalism of noiseless privacy to
define active and passive partial knowledge DP\@.

Other variants, which also model adversaries with partial background knowledge,
are not based on indistinguishability, but directly constrain the posterior
knowledge of an attacker as a function of their prior knowledge. Among those are
\emph{adversarial privacy}~\cite{rastogi2009relationship}, \emph{membership
privacy}~\cite{li2013membership}, and \emph{aposteriori noiseless
privacy}~\cite{bhaskar2011noiseless}.
\iffullversion%
It is straightforward to adapt the examples given in this paper to show that
these definitions suffer from the same flaws as noiseless privacy when data has
correlations. 
\fi
These definitions also also do not delineate between passive and active
attackers. Because of space constraints, we do not study them in detail.

Several other definitions have been proposed. \emph{Pufferfish
privacy}~\cite{kifer2012rigorous} can be seen as a generalization of noiseless
privacy, and similarly, \emph{coupled-worlds privacy}~\cite{bassily2013coupled}
(and its \emph{inference-based} variant) generalizes distributional differential
privacy: instead of protecting individual tuples, they protect arbitrary
sensitive properties of the data. It is straightforward to generalize our
results to the more generic frameworks.

\subsection{Background on existing definitions}

We first recall the original definitions of $\epsdel$-indistinguishability and
$\epsdel$-differential privacy.

\begin{definition}[$\epsdel$-indistinguishability~\cite{dwork2006calibrating}]
  Two random variables $A$ and $B$ are $\epsdel$-indistinguishable if for all
  measurable sets $X$ of possible events:
  \begin{align*}
    \proba{A\in X} & \leq e^{\eps}\cdot\proba{B\in X}+\del \\
    \proba{B\in X} & \leq e^{\eps}\cdot\proba{A\in X}+\del.
  \end{align*}
  We denote this as $A\indepsdel B$. If $\del=0$, we call this
  $\eps$-indistinguishability, and denote it by $\indeps$ (c.f.
  Table~\ref{table:notations}).
\end{definition}

In all the following, $\datasets$ designates the set of possible databases. A
database $D$ is a family of records: $D={\left(D(i)\right)}_{i\le n}$, where
each $D(i)$ is in a fixed set $\tuples$ of possible records, and $n$ is the size
of the set $D$. We only consider databases of fixed size $n$, and usually omit
the range of database indices $i$. Mechanisms, typically denoted $\mecha$, take
databases as input, and output some value in an output space $\outputs$. 

\begin{definition}[$\epsdel$-differential privacy~\cite{dwork2006calibrating}]\label{def:original}
  A privacy mechanism $\mecha$ satisfies \emph{$\epsdel$-differential privacy}
  (DP) if for any databases $D_1$ and $D_2$ that differ only on the data of one
  record, $\mech{D_1}\indepsdel\mech{D_2}$. If $\del=0$, we call this
  $\eps$-differential privacy.
\end{definition}

\iffullversion%
\fi%

\iffullversion%
\section{Correlated data}\label{sec:correlated}

When data is correlated, dependencies create problems for privacy definitions
that assume an attacker with partial knowledge. To illustrate this, we recall
two previously introduced definitions that model this situation differently:
\emph{noiseless privacy} (NP) and \emph{distributional differential privacy}
(DDP). We show that both definitions have undesirable consequences when data is
correlated, and use DDP as a starting point to solve this problem in two steps.
First, we modify DDP and introduce a new definition, \emph{causal differential
privacy} (CDP), to prevent its most direct problems. Second, we propose a
criterion that encompasses many use-cases but avoids known issues with
correlated data, and makes CDP equivalent to NP\@. This allows us to cleanly
define a rigorous notion of DP with partial knowledge, which allows for many
practical use cases, but avoids the known issues with correlated data.

Note that there has been substantial debates about the impact of correlations on
the guarantees that DP provides. The debates are summarized
in~\cite{tschantz2017differential}, where the authors suggest a possible
resolution: interpreting DP as a \emph{causal} property. In this section, we
continue this line of work in the context of partial knowledge. In particular,
we show that modifying the definition in the same way as the ``causal variants''
of~\cite{tschantz2017differential} is not sufficient to solve all issues created
by the presence of correlations in the data, when the attacker only has partial
knowledge.

For simplicity, in this section, we only consider the case with $\del=0$. We
re-introduce $\del>0$ in Section~\ref{sec:passiveactive}.

\subsection{Existing notions}

The assumption that the attacker lacks background knowledge can be represented
by considering the input data to be \emph{noisy}. This idea was first proposed
in~\cite{duan2009privacy}, and was formalized in~\cite{bhaskar2011noiseless} as
\emph{noiseless privacy}. Instead of comparing two databases that differ in only
one record, it uses a \emph{probability} distribution $\theta$, conditioned on
the value of one record $D(i)$: the randomness in $\theta$ captures the
attacker's uncertainty. This probability distribution generates not only a
dataset $D$ but also the attacker's partial knowledge $B$ (with values in some
space $\knowledges$). For brevity, we abbreviate $\Prob_{(D,B)\sim\theta}$ as
$\Prob_\theta$, abbreviate the observation $\dhd$ by $\hatD$, and the
observation $\bhb$ by $\hatB$; these notations as well as others used in this
paper are summarized in Table~\ref{table:notations}.

\begin{table}[t]
  \centering
  \begin{tabular}{ll}
    \toprule
    $\approx_{\eps,\del}$ & $(\eps, \del)$-indistinguishable \\
    $\tuples$ &Set of possible records \\
    $\mathcal{D}$ &Set of possible databases \\
    $D$ & Database (typically, a random variable) \\
    $D(i)$ & Single record in D \\
    \iffullversion%
    $D_{-i}$ & Database $D$ with $D(i)$ removed \\
    $D_{i\rightarrow b}$ & Database $D$ with $D(i)$ replaced by value $b$ \\ 
    \fi
    $n$ & Size of the database \\
    $\mathcal{B}$ & Set of auxiliary information/partial knowledge \\
    $B$ & Partial (or ``background'') knowledge \\
    $\theta$ & Probability distribution on $\datasets$ or $\datasets\times\knowledges$ \\
    $\Theta$ & Set of probability distributions \\
    $\Prob_{\theta}$ & Abbreviation of $\Prob_{(D,B)\sim \theta}$ \\
    $\hatB$ & Observation of $B$; abbreviation of $B=\hatB$ \\
    $\hatD$ & Observation of $D$; abbreviation of $D=\hatD$ \\
    $\mathcal{O}$ & Output space of mechanisms \\
    $\mecha$ & Mechanism \\
    $X\cond{E}$ & Random variable $X$ conditioned on event $E$ \\
    \bottomrule
  \end{tabular}
  \caption{Notations used in this paper.}\label{table:notations}
\end{table}

\begin{definition}[$\Theps$-noiseless privacy~\cite{bhaskar2011noiseless}]\label{def:np}
  Given a family $\Theta$ of probability distribution on $\dk$, a mechanism
  $\mecha$ is \emph{$\Theps$-noiseless private} (NP) if for all $\tit$, all
  $\hbik$, all indices $i$ and all $a,b\in\tuples$ such that
  $\probas{\theta}{\hatB,D(i)=a}\neq0$ and $\probas{\theta}{\hatB,D(i)=b}\neq0$
  (we call this condition ``$\hatB$ is \emph{compatible with} $D(i)=a$ and
  $D(i)=b$''):
  \begin{equation}
    \mech{D}\cond{\theta,\hatB,D(i)=a}\indeps\mech{D}\cond{\theta,\hatB,D(i)=b}.
  \end{equation}
  Here, the notation $\mech{D}\cond{\theta,\hatB,D(i)=a}$ refers to the random
  variable defined by $\mech{D}$, where $D\sim\theta$, conditioned on the event
  ``$B=\hatB$ and $D(i)=a$''.
\end{definition}

The original intuition behind DP states that changing one record must not change
the output too much. NP attempts to capture this intuition for an attacker with
partial knowledge, but Bassily et al.~\cite{bassily2013coupled} argue that this
definition is too strong. The following example illustrates their argument.

\begin{example}\label{ex:cw}
  Assume $\theta$ has a global parameter $\mu$ that is either +1 or -1 with equal
  probabilities, and outputs $n$ normally distributed records with mean $\mu$ and
  a small standard deviation. Releasing the average of the record values is not
  NP\@: for all indices $i$, $\mech{D}\cond{\theta,D(i)=1}$ will be close to $1$
  and $\mech{D}\cond{\theta,D(i)=-1}$ will be close to $-1$, so the two
  distributions are very distinguishable. This happens even though the impact of a
  single record in the database is low: once $\mu$ is fixed, the random choice of
  $i$ is unlikely to have a large effect on the global average.
\end{example}

This example shows a definitional problem. If the attacker previously knows
$\mu$, revealing $\mech{D}$ does not give much additional information on the
target $D(i)$: an attacker with \emph{less} initial knowledge is considered
\emph{more} powerful. We study this in detail in Section~\ref{sec:criterion}.

The authors propose an alternative definition to fix this problem:
\emph{$\Theps$-distributional differential privacy}. It requires that $\mecha$
can be \emph{simulated} by another mechanism $\Simu$, that does not have access
to the sensitive property. The intuition is as follows: if $\mech{D}$ is close
to $\Sim{D_{-i}}$ for some simulator $\Simu$, then $\mecha$ cannot leak ``too
much'' about the value of $D(i)$.

\begin{definition}[$\Theps$-distributional differential privacy~\cite{bassily2013coupled}]\label{def:ddp}
  Given a family $\Theta$ of probability distributions on $\dk$, a mechanism
  $\mecha$ satisfies \emph{$\Theps$-distributional differential privacy}
  ($\Theps$-DDP) if there is a simulator $\Simu$ such as for all probability
  distributions $\tit$, all $\hbik$, all $i$, and all $a\in\tuples$ such that
  $\hatB$ is compatible with $D(i)=a$:
  \begin{equation*}
      \mech{D}\cond{\theta,\hatB,D(i)=a}\ind_\eps\Sim{D_{-i}}\cond{\theta,\hatB,D(i)=a},
  \end{equation*}
  where $D_{-i}$ is the database $D$ from which the record $i$ has been removed.
\end{definition}

The distribution $\theta$ and the mechanism $\mecha=\text{avg}$ from
Example~\ref{ex:cw} satisfy this definition: the simulator can be defined as
simply running $\mecha$ on $D_{-i}$, possibly after adding $+1$ or $-1$
depending on the other records.

\subsection{Distributional differential privacy under correlations}

This critique of NP is similar to the critique of the associative view of DP
in~\cite{tschantz2017differential}. But the proposed fix has a flaw: $\Simu$ can
use strong dependencies in the data to artificially satisfy the definition. In
the following example, the values of different records are strongly correlated,
and the simulator \emph{cheats} by using these correlations: consequently, the
identity function is considered private!

\begin{example}\label{ex:duplicates}
  Let $\theta$ output $n$ \emph{duplicate} records: for all $i<n$, $D(2i)$ is
  picked from some probability distribution $R$, and $D(2i+1)=D(2i)$. Then the
  identity function $\text{Id}$, which simply outputs its input without any
  noise, is $(\{\theta\},0)$-DDP\@! Indeed, the simulator can simply replace
  the missing record by its duplicate and output the entire database:
  $\text{Id}(D)$ is exactly the same as $\Sim{D_{-i}}$.
\end{example}

Here, the dependency relationships are ``extreme'', as each record is
duplicated. But even when records are less strongly correlated, the problem is
still present. In fact, the more dependencies are in the data, the more
accurately the simulator can simulate the missing record, and the more
``private'' the mechanism is (since $\eps$ gets lower): a more powerful
adversary, who can exploit dependencies in the data, is considered weaker by the
definition. This is clearly undesirable.

How can we formalize an adversary that cannot ``cheat'' using dependencies in
the data? We propose one possible option: using the same technique as the causal
variants of DP described in~\cite{tschantz2017differential}, we simply change
the target record \emph{after} the distribution is generated.

\begin{definition}[$\Theps$-causal differential privacy]\label{def:cdp}
  Given a family $\Theta$ of probability distributions on $\dk$, a mechanism
  $\mecha$ satisfies \emph{$\Theps$-causal differential privacy} ($\Theps$-CDP)
  if for all probability distributions $\theta\in\Theta$, all $i$, all
  $a,b\in\tuples$, and all $\hbik$ compatible with $D(i)=a$ and $D(i)=b$:
  \begin{equation*}
      \mech{D}\cond{\theta,\hatB,D(i)=a}\ind_\eps\mech{\Darrow}\cond{\theta,\hatB,D(i)=a},
  \end{equation*}
  where $\Darrow$ is the database $D$, where the $i$-th record has been replaced
  by $b$.
\end{definition}

CDP still captures DP's intuition: the change in one data point should not
influence the output of the mechanism too much. However, the change happens
\emph{after} the influence of the dependencies in the data. This version is
strictly stronger than the original version: if a mechanism $\mecha$ is
$\Theps$-CDP, then it is also $\Theps$-DDP\@. Indeed, the simulator $\Simu$ can
always replace the missing record with an arbitrary value $b$ and return
$\mech{\Darrow}$.

In~\cite{bassily2013coupled}, the authors also introduce an
\emph{inference-based} version of DDP\@. We can as easily adapt CDP to this
different formalization.

\begin{definition}[$\Theps$-inference-based causal differential privacy]\label{def:ibcdp}
  Given a family $\Theta$ of probability distributions on $\dk$, a mechanism
  $\mecha$ satisfies \emph{$\Theps$-inference-based causal differential privacy}
  ($\Theps$-IBCDP) if for all probability distributions $\tit$, for all indices
  $i$, all $b\in\tuples$, all $t\in\outputs$, and all $\hbik$ compatible with
  $\mech{D}=t$ and $\mech{\Darrow}=t$:
  \begin{equation*}
    {D(i)}\cond{\theta,\hatB,\mech{D}=t}\ind_\eps{D(i)}\cond{\theta,\hatB,\mech{\Darrow}=t}
  \end{equation*}
  where $\Darrow$ is the database $D$, where the record $i$ has been replaced by
  $b$.
\end{definition}

Note that this definition is equivalent to the indistinguishability-based
version (Definition~\ref{def:cdp}) up to a change in parameters.

\begin{prop}\label{prop:cdp-ibcdp}.
  $\Theps$-CDP implies $\Theeps$-IBCDP, and $\Theps$-IBCDP implies
  $\Theeps$-CDP\@.
  \begin{proof}\label{proof:cdp-ibcdp}
  The first implication can be proven in the same way as Theorem~1
  in~\cite{bassily2013coupled}, replacing $\Sim{D_{-i}}$ by $\mech{\Darrow}$. For
  the second implication, suppose that a mechanism $\mecha$ is $\Theps$-IBCDP, and
  assume that the attacker has no background knowledge. Consider an index $i$, two
  possible record values $a,b\in\tuples$, and one possible output value
  $t\in\outputs$. Bayes' rule gives us:
  \begin{align*}
    \frac{\proba{\mech{\Darrow}=t \vert D(i)=a}}{\proba{\mech{D}=t \vert D(i)=a}}
    = \frac{\proba{D(i)=a\vert\mech{\Darrow}=t}}{\proba{D(i)=a\vert\mech{D}=t}}
      \cdot \frac{\proba{\mech{D}=t}}{\proba{\mech{\Darrow}=t}}.
  \end{align*}
  The first term is between $e^{-\eps}$ and $e^{\eps}$ since $\mecha$ is
  $\Theps$-IBCDP\@. We only need to show that the second term is also between
  $e^{-\eps}$ and $e^{\eps}$ to conclude the proof.
  Notice that when $D(i)=b$, we have $\Darrow=D$. Thus:
  \begin{align*}
    1 = \frac{\proba{\mech{\Darrow}=t \vert D(i)=b}}{\proba{\mech{D}=t \vert D(i)=b}}
      = \frac{\proba{D(i)=b\vert\mech{\Darrow}=t}}{\proba{D(i)=b\vert\mech{D}=t}}
        \cdot \frac{\proba{\mech{\Darrow}=t}}{\proba{\mech{D}=t}}.
  \end{align*}
  Again, the first term is between $e^{-\eps}$ and $e^{\eps}$ since $\mecha$ is
  $\Theps$-IBCDP\@. Since multiplying it with the second term gives $1$, the
  second term is also between $e^{-\eps}$ and $e^{\eps}$. If the attacker has
  background knowledge, all the probabilities above are conditioned by $\hatB$, and
  the same reasoning holds.

  In the more general case where the attacker does have some partial knowledge,
  all probabilities above are conditioned by the value of this partial knowledge,
  and the same reasoning holds.
  \end{proof}
\end{prop}

This equivalence is only true in the context of this section, where $\del=0$; we
explain later why it fails when $\del>0$.

Does Example~\ref{ex:cw} satisfy CDP\@? It depends: if $b$ can take
arbitrarily large values, $\text{avg}\left(D_{i\rightarrow b}\right)$ can be
arbitrarily distinguishable from $\text{avg}(D)$. Otherwise, $b$ can only have a
bounded influence on the average and $\theps$-CDP can hold for some $\eps$.
In other words, when using the fixed version of the definition, whether a
given mechanism is CDP depends on the \emph{sensitivity} of the mechanism. This
is a good thing: it suggests that it captures the same intuition as DP\@.

Example~\ref{ex:cw} shows that CDP is not stronger than NP\@. Is the reverse
true? In Example~\ref{ex:outliers}, we show that this is not the case.

\begin{example}\label{ex:outliers}
  Consider the same $\theta$ as for Example~\ref{ex:cw}: it depends on a global
  parameter, $\mu$, which is either +1 or -1 with equal probabilities, and each
  of the $n$ records is normally distributed with mean $\mu$ and a small
  standard deviation $\sigma$. Let $\mecha$ be the algorithm that counts
  \emph{outliers}: it computes the average $\tilde{\mu}$ of all data points, and
  returns the number of records outside
  $\left[\tilde{\mu}-5\sigma,\tilde{\mu}+5\sigma\right]$. As we saw before,
  conditioning $\theta$ on a value of $D(i)$ is approximately equivalent to
  fixing $\mu$: the number of outliers is going to be the same no matter what (0
  with high probability). However, if we first condition $\theta$ on $D(i)=a$,
  and then change this record into $b$, we can choose $b$ so that this record
  becomes an outlier; and make it 1 with high probability. Thus, this mechanism
  is NP but not CDP\@.
\end{example}

Even though Example~\ref{ex:outliers} shows that NP does not imply CDP, it is
natural to think that in many cases, if you change one data point $i$ \emph{as
well as all data points correlated with it}, it will have a bigger influence on
the algorithm that if you only change $i$ without modifying the rest of the
data. In Example~\ref{ex:duplicates-integers}, we show that even for a simple
data dependencies and mechanisms, we can find counterexamples to this intuition.

\begin{example}\label{ex:duplicates-integers}
  Consider a probability distribution $\theta$ that outputs $2n$ records, such as
  for all $i<n$, $D(2i)$ is picked from some probability distribution $R$ with
  values in $\mathbb{N}$, and $D(2i+1)=D(2i)$. Then the mechanism that sums all
  records might be in NP, but cannot be in CDP\@. Indeed, if $D\sim\theta$, then
  $\sum_i D(i)$ will always be even, but changing one record without modifying
  its duplicate can make the sum odd.
\end{example}

This last example that finding a special case where NP implies CDP is likely
difficult. There is, however, a special case where both are equivalent: the
absence of dependencies in the data. If changing one record does not influence
other records, then NP and CDP are equivalent. This result is similar to
Corollary~2 in~\cite{bassily2013coupled}, but is simpler and without the change
in parameters.

\begin{prop}\label{prop:np-cdp-independence}
  Let $\Theta$ be a family of probability distributions such that for all $\tit$
  and all $\hbik$, the random variables ${D(i)}\cond{\theta,\hatB}$ are mutually
  independent. Then a mechanism $\mecha$ is $\Theps$-NP iff it is
  $\Theps$-CDP\@.
  \begin{proof}
    Under these conditions, ${D_{-i}}\cond{\theta,\hatB,D(i)=a}$ is exactly
    ${D_{-i}}\cond{\theta,\hatB,D(i)=b}$, so
    $\mech{D}\cond{\theta,\hatB,D(i)=b}$ is the same as
    $\mech{\Darrow}\cond{\theta,\hatB,D(i)=a}$. The statement follows.
  \end{proof}
\end{prop}

This natural property, combined with the better behavior of CDP in scenarios
like Example~\ref{ex:duplicates}, might seem like CDP is a better alternative to
CDP, when one wants to capture an attacker with partial knowledge, under the
causal interpretation of differential privacy. However, even with this fix, when
records are not independent, CDP is not always safe to use. We present an
example from Adam Smith (personal correspondence, 2018-09-28) showing that a 
slightly modified version of the identity function can still be CDP if records
are strongly correlated.

\begin{example}\label{ex:triplicates}
  Let $\theta$ output $3n$ \emph{triplicated} records: for all $i<n$, $D_{3i}$
  is picked from some probability distribution $R$, and $D(3i+1)=D(3i+2)=D(3i)$.
  Let $\mecha$ be a mechanism that ``corrects'' a modified record: if there is a
  record value $x$ appearing only once, and a value $y$ appearing only twice,
  then $\mecha$ changes the record $x$ to $y$; then $\mecha$ always outputs the
  entire database. It is easy to check that $\mecha$ is $(\{\theta\},0)$-CDP\@.
\end{example}

This example is more artificial than Example~\ref{ex:duplicates}, as the
mechanism itself ``cheats'' to use dependencies in the data. Nonetheless, it
shows that some mechanisms that leak the full database can be CDP\@. Thus, using
CDP as a privacy measure of a given mechanism is dangerous if no information
about the mechanism is known. We do not know whether more natural mechanisms
could lead to similar counterexamples, for certain classes of probability
distributions; but it is clear that simply applying the same technique as the
causal variants of~\cite{tschantz2017differential} is insufficient to solve
entirely the problems with correlations under partial knowledge.

\subsection{Imposing an additional criterion on the definition}\label{sec:criterion}

In this section, we propose a criterion that the distribution $\theta$ must
satisfy before NP can be used. We argue that when this criterion is not
satisfied, NP is not a good measure of the privacy of a mechanism. But when it
is satisfied, we obtain natural properties that are false in general: NP and CDP
are equivalent, and attackers with more partial knowledge are stronger.

We mentioned previously that NP was not \emph{monotonous}: an attacker with
\emph{more} knowledge can be considered to be \emph{less powerful}. In
Example~\ref{ex:cw}, an attacker $A$ who did not know $\mu$ can \emph{learn}
$\mu$ by observing $\mech{D}$. This increases her knowledge about $D(i)$: she
now knows that $D(i)$ is probably around $\mu$, a fact previously unknown.
However, an attacker $B$, who \emph{already knew} $\mu$, does not increase her
knowledge as much when observing $\mech{D}$. Examples~\ref{ex:duplicates}
and~\ref{ex:triplicates} show that DDP and CDP also suffer from this issue. How
can we fix this problem? The informal goal is that an attacker with more
background knowledge should gain more information. We must make sure that the
privacy quantification $\eps$ cannot be artificially inflated by non-sensitive
information learned by the attacker. 

In all previous examples, the attacker's partial knowledge is strongly
correlated with the sensitive information. Thus, $\eps$ does not measure the
privacy loss \emph{due to the mechanism}, but also takes into account the prior
knowledge from the attacker about the sensitive attribute. Modeling the
attacker's uncertainty is necessary to formalize her partial knowledge, but the
only thing that should be captured by $\eps$ is the \emph{mechanism}'s privacy
leakage. To ensure this is the case, we argue that the partial knowledge must be
independent from the sensitive information, and we propose a formalization that
enforces this distinction between sensitive information and partial knowledge.


To this end, we propose an alternative way to model the attacker's uncertainty,
and suggest to \emph{normalize} the distribution $\theta$ to cleanly separate
sensitive information and partial knowledge. We show that if such a
normalization exists, then an attacker with more partial knowledge is more
powerful. Thus, the existence of such a normalization is a desirable property
for privacy definitions that model an attacker with partial knowledge, and we
argue that it should serve as a \emph{criterion} that must be satisfied before
using such definitions, in order to get meaningful results.

How to formalize the intuition that the sensitive information should be
separated from the partial knowledge? The core idea is to express $\theta$ as
the output of a \emph{generative function}, with independent random parameters.
Each possible value of these parameters corresponds to a possible database. 

\begin{definition}[Normalization of data-generating distributions]\label{def:normalization}
  A \emph{normalization} of a probability distribution $\theta$ is a family of
  mutually independent random variables $\left(\phi_0,\dotsc,\phi_k\right)$, and
  an injective, deterministic function $\hatheta$, such that
  $\theta=\hatheta\left(\phi_0,\dotsc,\phi_k\right)$.
\end{definition}

A normalization \emph{de-correlates} the distribution: it splits its randomness
into independent parts $\phi_i$. The $\phi_i$ can then play distinct roles: one
parameter can capture the sensitive property, while the others can model the
attacker's partial knowledge. We capture this additional requirement in the
following definition.

\begin{definition}[Acceptable parameters]\label{def:acceptable}
  Given a distribution $\theta$ with values in $\dk$, an \emph{acceptable
  normalization} of $\theta$ at index $i$ is a normalization
  $\hatheta\left(\phi_0,\dotsc,\phi_k\right)$ where:
  \begin{enumerate}
    \item $\phi_0$ entirely determines the sensitive attribute $D(i)$: there
      exists a function $f$ such that for all possible values of $D(i)$,
      $\probas{\theta}{f\left(\phi_0\right)=D(i)}=1$.
    \item Some of the $\phi_1,\dotsc,\phi_k$ entirely determine the partial
      knowledge: there exists $I\subseteq\{1,\dotsc,k\}$ and an injective
      function $g$ such that
      $\probas{\theta}{g\left({\left(\phi_j\right)}_{j\in I}\right)=B}=1$.
  \end{enumerate}
  A family of distributions $\Theta$ is \emph{acceptable} if each $\tit$ has an
  acceptable normalization at all indices $i$.
\end{definition}

In practice, we can simply consider a family ${\left(\phi_j\right)}_{j\in I}$
for some $I$ to be the attacker's partial knowledge, rather than using a
bijection. When partial knowledge is defined using a subset of parameters, an
attacker has more partial knowledge when she knows more parameters.

Informally, $\phi_0$ must contain enough information to retrieve the sensitive
attribute, and the partial knowledge must be independent from it. How can we
normalize the $\theta$ in Example~\ref{ex:cw}? We cannot have one parameter for
$\mu$, and $n$ parameters for the noise added to each record: the sensitive
attribute $D(i)$ would require two parameters to express. Rather, $\phi_0$ could
be a pair containing \emph{both} $\mu$ and the noise added at $i$, $D(i)-\mu$.
The attacker's partial knowledge can be $D(j)-\mu$, for $j\neq i$, without
$\mu$. In other words, $\mu$ itself is also sensitive. What if we do not want to
consider $\mu$ as sensitive? Then, $\phi_0$ can be the value of the noise added
to the record $i$, $D(i)-\mu$, and $\mu$ is another parameter that can (or not)
be part of the attacker's partial knowledge.

As this example shows, this separation between the partial knowledge and the
sensitive value forces us to carefully choose the sensitive value, and it
prevents us from comparing scenarios where the sensitive value varies. This
formalism can now be used to precisely define the relative strength of two
attackers, based on their partial knowledge, and show that an attacker with more
knowledge is more powerful.

\begin{definition}[Relative strength of partial knowledge]
  Given probability distributions $\theta_1$ and $\theta_2$ with values in
  $\dk$, we say that $\theta_1$ has \emph{more background knowledge} than
  $\theta_2$ if the three following conditions are satisfied.
  \begin{enumerate}
    \item $D\cond{\theta_1}=D\cond{\theta_2}$: the only difference
      between the probability distributions is the partial knowledge.
    \item For all $i$, there exists an acceptable normalization
      $\hatheta\left(\phi_0,\dotsc,\phi_k\right)$ given $i$ that is common to
       $\theta_1$ and $\theta_2$.
    \item For all $i$, if we denote $I_1$ and $I_2$ the set of parameters that
      correspond to $\theta_1$ and $\theta_2$ respectively in this acceptable
      normalization, then $I_1 \supseteq I_2$.
  \end{enumerate}
  Given two families of probability distributions $\Theta_1$ and $\Theta_2$, we
  say that $\Theta_1$ has more background knowledge than $\Theta_2$ if for all
  $\theta_2\in\Theta_2$, there exists $\theta_1\in\Theta_1$ such that $\theta_1$
  has more background knowledge than $\theta_2$.
\end{definition}

\begin{prop}\label{prop:more-knowledge}
  Let $\Theta_1$ and $\Theta_2$ be two families of distributions such that
  $\Theta_1$ has more background knowledge than $\Theta_2$. If a mechanism
  $\mecha$ satisfies $\left(\Theta_1,\eps\right)$-NP, it
  also satisfies $\left(\Theta_2,\eps\right)$-NP\@.
  \begin{proof}
    Suppose that $\mecha$ is $\left(\Theta_1,\eps\right)$-NP\@. For a
    distribution $\theta_2\in\Theta_2$ and an index $i$, let
    $\theta_1\in\Theta_1$ be such that $\theta_1$ is stronger than $\theta_2$.
    By definition, there exists an acceptable normalization
    $\hatheta\left(\phi_0,\dots,\phi_k\right)$ common to $\theta_1$ and
    $\theta_2$. Let $f$ be the function extracting the sensitive value from
    $\phi_0$ in this normalization. Denoting $I_1$ and $I_2$ the set of
    parameters corresponding respectively to $\theta_1$ and $\theta_2$, as a
    simplification, we assume that $I_2 = \varnothing$ and $I_1 = \{1\}$; it is
    straightforward to adapt the proof to the more generic case. For any output
    $O$, and all values $a,b\in\tuples$, we can decompose:
    \begin{align*}
      \probac{\mech{D}=O}{f(\phi_0)=a}
        = \sum_{\hatB} \probac{\phi_1=\hatB}{f(\phi_0)=a}
            \cdot \probac{\mech{D}=O}{f(\phi_0)=a,\phi_1=\hatB}.
    \end{align*}
    The $\phi_i$ are independent: $\probac{\phi_1=\hatB}{f(\phi_0)=a}$ is
    the same as $\probac{\phi_1=\hatB}{f(\phi_0)=b}$. Since $\mecha$ satisfies
    $\left(\Theta_1,\eps\right)$-NP, we also have for all $\hatB$:
    \begin{align*}
      \probac{\mech{D}=O}{f(\phi_0)=a,\phi_1=\hatB}
        \le e^\eps\probac{\mech{D}=O}{f(\phi_0)=b,\phi_1=\hatB}.
    \end{align*}
    Thus:
    \begin{align*}
      \probac{\mech{D}=O}{f(\phi_0)=a} 
        & \le e^\eps \sum_{\hatB} \probac{\phi_1=\hatB}{f(\phi_0)=b} 
            \cdot \probac{\mech{D}=O}{f(\phi_0)=b\wedge\phi_1=\hatB} \\
        & \le e^\eps \probac{\mech{D}=O}{f(\phi_0)=b} 
    \end{align*}
    and thus, $\mecha$ satisfies $\left(\Theta_2,\eps\right)$-NP\@.
    Adapting the proof to cases where $I_1 \subseteq I_2$ is straightforward.
  \end{proof}
\end{prop}

This proposition states that, for acceptable distributions, partial background
knowledge can be formalized in a reasonable and intuitive way, guaranteeing that
attackers with more background knowledge are stronger. Another advantage is that
when this criterion holds, NP and CDP are equivalent.

\begin{prop}\label{prop:np-cdp-acceptable}
  If $\Theta$ is an acceptable distribution, then for any $\eps$ and $\del$,
  $\Thepsdel$-NP is equivalent to $\Thepsdel$-CDP\@.
\end{prop}

The proof is the same as for Proposition~\ref{prop:np-cdp-independence}.
Figure~\ref{fig:relation-np-ddp-cdp} summarizes the relations between the
definitions introduced in this section.

\begin{figure}[!h]
    \centering
    \includegraphics[scale=0.8]{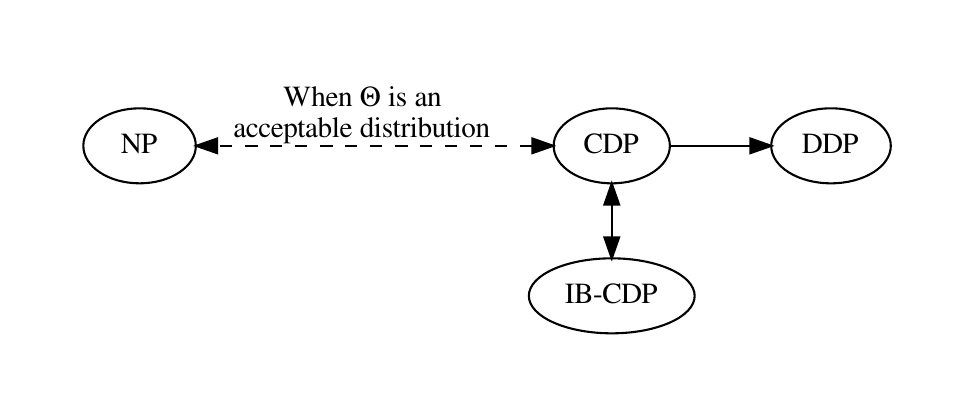}
    \caption{Relations between definitions introduced in
    Section~\ref{sec:correlated}, assuming
    $\del=0$.}\label{fig:relation-np-ddp-cdp}
\end{figure}

Acceptable normalizations force practitioners to define which information is
considered private. If, like in DP, the sensitive information is the value of a
given record, with an acceptable normalization, the attacker's partial knowledge
cannot contain records correlated with the target record. This might seem overly
restrictive: what if the attacker \emph{does} know some information correlated
with the target record? In this case, one must change the sensitive information
to only consider the decorrelated part as sensitive.

To illustrate this process, consider a medical database where the diagnostic
records of different patients might be correlated. To find an acceptable
normalization, we must choose between two options. The first is to consider
attackers that do not have information correlated with the diagnosis of a given
patient: in that case, we must protect the information of multiple patients at
once, similarly to \emph{DP under correlations} (a variant of DP defined
in~\cite{chen2014correlated}). Another option is for the sensitive property to
be the diagnosis of someone \emph{given the diagnosis of those they are
correlated with}. We formally present a simpler example below, in
Exemple~\ref{ex:referendum-normalization}.

\begin{example}\label{ex:referendum-normalization}
  Consider a referendum, where people vote in pairs, with some amount of
  correlation between pairs. More precisely, $\theta$ is a distribution that
  generates a database of $2n$ records according to the following process:
    \begin{itemize}
      \item $D(2i)=1$ with probability $p_i$, and $0$ with probability $1-p_i$;
      \item $D(2i+1)=D(2i)$ with probability $p_c$, and $1-D(2i)$ with probability
        $1-p_c$.
    \end{itemize}
  Suppose the attacker is interested in $D(1)$. There are two ways of modeling
  this with an acceptable normalization, depending on whether we want to allow
  the attacker to know $D(0)$.
    \begin{itemize}
      \item We can pick $\phi_{i}$ to determine the value of \emph{both} $D(2i)$
        and $D(2i+1)$. This way, $\phi_0$ is sufficient to know the value of the
        sensitive information $D(i)$, and is independent from all $\phi_i$ for
        $i>0$.
      \item We can also pick $\phi_{2i}$ to be the event ``$D(2i)=D(2i+1)$'', and
        $\phi_{2i+1}$ to be the value of $D(2i)$. In that case, the attacker is
        allowed to know $D(0)$, but the sensitive value has changed: the
        sensitive value is whether $D(0)=D(1)$; which is independent from the
        actual value of $D(0)$.
    \end{itemize}
\end{example}

\fi%

\section{Passive vs.\ active attackers}\label{sec:passiveactive}

\iffullversion%
In this section and the rest of this paper, we assume that the criterion
introduced in Definition~\ref{def:acceptable} is satisfied: the data-generating
distributions considered are always acceptable. For simplicity, we also assume
that the sensitive property is the value of one record. Under these conditions,
any value of the partial knowledge $\hatB$ is compatible with any value of the
sensitive record $D(i)=a$, since these two events are independent. This also
allows us to only consider the formalism of NP\@; since it is equivalent with
CDP under this criterion (Proposition~\ref{prop:np-cdp-acceptable}).

Correlations in the data are not the only issue to account for when limiting the
attacker's background knowledge in DP\@. Another 
\fi%
important question is whether the attacker simply \emph{receives} partial
knowledge passively, or whether the attacker can \emph{actively influence} the
data. In this section, we show how to model these two situations by adapting the
notion of a privacy loss random variable to model an attacker with partial
knowledge, and we explore the relationship between the two corresponding
definitions. To help understand this distinction, in this section, we consider
the following example.

\begin{example}[Thresholding]\label{ex:thresholding}
  1000 people take part in a Yes/No-referendum. Each person votes ``Yes'' with
  some probability, independently from the others. The mechanism $\mecha$ counts
  the number of ``Yes'' votes, but only returns this if it is above 100;
  otherwise it returns $\bot$. The partial knowledge $B$ contains the votes of
  100 participants, and the attacker wants to know the vote of another
  individual $D(i)$. We will see that in case the probability that each person
  votes ``Yes'' is very small (say, $10^{-7}$), the privacy of this scheme will
  depend on whether the attacker is passive or active.
\end{example}

\subsection{Privacy loss random variable}

Let us recall the \emph{privacy loss random variable}~\cite{dinur2003revealing}
(PLRV). For simplicity, we only consider the case where the set of possible
outputs of the mechanism, $\outputs = \bigcup_{D}\supp(M(D))$, is countable.

\begin{definition}[Privacy Loss Random Variable~\cite{dinur2003revealing}]\label{def:plrv}
  Given a mechanism $\mecha$, and two datasets $D_1$ and $D_2$, the privacy loss
  random variable (PLRV) of an output $O\in\outputs$ is defined as
  \begin{equation*}
    \privlossM{D_1}{D_2}(O)=\ln\frac{\proba{\mech{D_1}=O}}{\proba{\mech{D_2}=O}}.
  \end{equation*}
  Using the convention $x/0=\infty$ for all $x$, the PLRV can be $\pm\infty$.
\end{definition}

$\epsdel$-DP can be reformulated using the PLRV\@.

\begin{lemma}[{\cite[Lemma 1]{meiser2018tight}}]\label{lemma:dp-with-plrv}
  A mechanism $\mecha$ is $(\eps,\del)$-DP iff for all neighboring databases
  $D_1,D_2\in\datasets$ (differing in $1$ record):
  \begin{equation*}
    \expectu{O\sim\mech{D_1}}{\max(0,1-e^{\eps-\privloss{D_1}{D_2}(O)})} \le \del.
  \end{equation*}
\end{lemma}

Suppose the attacker only has partial knowledge about the data: the data comes
from a distribution $\theta$, and the attacker tries to distinguish between
$D(i)=a$ and $D(i)=b$ by observing $\mech{D}$, given partial knowledge $\hatB$.
Since $\hatB$ is given to the attacker \emph{prior} to $\mech{D}$, we must
\emph{condition} the probabilities by $\hatB$.

\begin{definition}[PLRV for partial knowledge]\label{def:plrv-pk}
  Given a mechanism $\mecha$, a distribution $\theta$ with values in $\dk$, an
  index $i$, and values $a,b\in\tuples$, the PLRV of an output $O\in\outputs$
  given partial knowledge $\hatB$ is:
  \begin{equation*}
  \privlossMit{a}{b}(O,\hatB) =
    \ln\frac{
      \probasc{\theta}{\mech{D}=O}{D(i)=a,\hatB}
    }{
      \probasc{\theta}{\mech{D}=O}{D(i)=b,\hatB}
    },
  \end{equation*}
  using the convention $x/0=\infty$ for all $x$.
\end{definition}

This PLRV captures the same idea as for classical DP\@: it quantifies the
attacker's information gain. If $B=D_{-i}$, this definition is the same as the
classical PLRV\@.

Now that we translated the concept of PLRV to account for partial knowledge, we
can use it to adapt the privacy definition. The formula in
Lemma~\ref{lemma:dp-with-plrv} averages the PLRV over all possible outputs $O$,
but the PLRV with partial knowledge has a second parameter, $\hatB$. How should
this new parameter be handled? There are at least two reasonable possibilities.

\subsection{Active partial knowledge}

The first option is to quantify over \emph{all} possibilities for the attacker's
partial knowledge. We assume the worst: we consider the case where the
attacker's partial knowledge causes the privacy to be the greatest. This models
a scenario where the attacker can not only see, but also \emph{influence} the
data. If the attacker can, for example, add fake users to the database, then she
can choose the values associated to these new records to maximize the chances of
information gain. We therefore call this option \emph{active} partial knowledge,
short for ``partial knowledge under active attacks''.


\begin{definition}[APKDP]\label{def:apk-dp}
  Given a family of distributions $\Theta$, a mechanism $\mecha$ is
  $\Thepsdel$-APKDP (Active Partial Knowledge Differential Privacy) if for all
  distributions $\tit$, all indices $i$, all $a,b\in\tuples$, and all $\hbik$:
  \begin{equation*}
    \expectu{\theta\cond{D(i)=a,\hatB},O\sim\mech{D}}{\max\left(0,1-e^{\eps-\privlossMit{a}{b}(O,\hatB)}\right)}\le\del,
  \end{equation*}
  or, equivalently:
  \begin{equation*}
    \mech{D}\cond{\theta,D(i)=a,\hatB}\ind_\epsdel\mech{D}\cond{\theta,D(i)=b,\hatB}.
  \end{equation*}
\end{definition}

The proof of this equivalence is the same as in~\cite[Lemma~1]{meiser2018tight},
which makes it explicit that APKDP is the same as NP in its reformulation
in~\cite{bassily2013coupled}. As shown in~\cite{tschantz2017differential},
APKDP and DP coincide whenever the attacker has full knowledge (when
$B=D_{-i}$).

With APKDP, a fixed part of the distribution can be arbitrarily determined. In
Example~\ref{ex:thresholding}, this corresponds to the attacker controlling
some percentage of voters. Such an active attacker can simply add many fake
``Yes'' votes to the database to reach the threshold of 100, rendering the
thresholding pointless. $\mecha$ then becomes a simple counting query without
providing privacy. With high probability, everybody votes ``No'', and the only
uncertainty left is over the attacker's target.

In addition to modeling an active attacker, APKDP can also be used in scenarios
where $\theta$ is unknown, but can be \emph{approximated}. The ``partial
knowledge'' can represent the error between the true distribution and the
approximation and if APKDP is satisfied, then the privacy property also holds
for the true database.

Note that in this context, explicitly modeling the background knowledge $\hatB$
is technically not necessary. Instead, we could simply create a new family of
probability distributions $\Theta'$ by conditioning each $\tit$ by the value of
each possible $\hatB$. We make this background knowledge explicit instead, so
APKDP is easier to compare with PPKDP, defined in the next section.

\subsection{Passive partial knowledge}

APKDP represents situations where the attacker can modify the data. An example
is an online service that publishes statistics about its use, where the attacker
can interact with the service before the usage statistics are published. Now,
what if the attacker cannot interact with the data? Consider e.g.\ researchers
publishing the results of a clinical study about patients having a medical
condition. A typical attacker cannot influence the clinical data, but might have
some partial knowledge about other participants to the survey.

How can we model such a \emph{passive} attacker, which has access to some
information about the data, but cannot influence it? It no longer makes sense to
quantify over arbitrary partial knowledge. In the same way that the
reformulation of $\epsdel$-DP using the PLRV averages the PLRV over all possible
outputs, we must average the PLRV over all possible values of the partial
knowledge.

\begin{definition}[PPKDP]
  Given a family of distributions $\Theta$, a mechanism $\mecha$ is
  $\Thepsdel$-PPKDP (Passive Partial Knowledge Differential Privacy) if for all
  distributions $\tit$, all indices $i$, and all $a,b\in\tuples$:
  \begin{equation*}
    \expectu{\theta\cond{D(i)=a},O\sim\mech{D}}{\max\left(0,1-e^{\eps-\privlossMit{a}{b}(O,B)}\right)}\le\del.
  \end{equation*}
\end{definition}

In this context, $\del$ has a similar meaning as in $\epsdel$-DP\@: it captures
the probability that the attacker is \emph{lucky}. In $\epsdel$-DP, it means
that $O$ allows the attacker to distinguish between $D_1$ and $D_2$ with high
probability, or, equivalently, that the PLRV associated to output $O$ is large.
In $\epsdel$-PPKDP however, $\del$ captures the probability of the attacker
getting either a favorable output $O$, or a favorable partial knowledge $B$.

With PPKDP, the thresholding mechanism of Example~\ref{ex:thresholding} is
private. Indeed, with high probability, the partial knowledge will have only
``No'' votes; and almost certainly, the mechanism will output $\bot$ and gives
no information. We formalize this intuition in Section~\ref{sec:thresholding}.

Note that as opposed to APKDP, PPKDP cannot be easily reformulated using
$\epsdel$-indistinguishability. Since the statement $\bhb$ \emph{conditions}
both probabilities, the $\del$ in $\epsdel$-indistinguishability only applies to
the randomness of $\mecha$. To use an indistinguishability-based formulation, we
would need to use $\del$ twice, and (for example) explicitly require that
$\epsdel$-indistinguishability holds with probability $1-\del$ over the choice
of $\hatB$.

\begin{remark}
  PPKDP shares some characteristics with inference-based distributional
  differential privacy (IBDDP), introduced in~\cite{bassily2013coupled}. A
  mechanism $\mecha$ satisfies IBDDP if there is a simulator $\Simu$ such that
  for all probability distributions $\tit$, and indices $i$\iffullversion, 
  the statement:\fi
  \begin{align*}
    {D(i)}\cond{\theta,\mech{D}=\hatO,\hatB}\ind_\epsdel{D(i)}\cond{\theta,\Sim{D_{-i}}=\hatO,\hatB}
  \end{align*}
  holds with probability $1-\del$ over the choice of $\hatO$ and $\hatB$.

  Leaving aside the simulator, note that $\del$ is used in \emph{two} separate
  parts of the definition: both over the choice of $\hatO$ and $\hatB$, and in
  the indistinguishability. As such, it is difficult see intuitively what $\del$
  corresponds to, and the interpretation based on the ``probability that the
  attacker gets more information than $e^\eps$'' is not correct. This is one of
  the reasons why a PLRV-based formulation is more convenient: $\del$ can simply
  be interpreted in the same way as in $\epsdel$-differential privacy.

  Further, the strict implication between DDP and IBDDP proven
  in~\cite{bassily2013coupled} can be explained by a similar distinction between
  an active and a passive attacker, even if it is not made explicit in the
  original paper. Indeed, when $\del>0$, this $\del$ applies only to the
  indistinguishability property of DDP, and DDP quantifies over all possible
  values of the background knowledge: the attacker is assumed to be able to
  choose the most favorable value of the background knowledge. In contrast, with
  IBDDP, $\del$ is applied both to the indistinguishability property \emph{and}
  to the choice of the background knowledge; hence the attacker is implicitly
  assumed to get the background knowledge randomly.
\end{remark}
\vspace{-1em}

\subsection{Relation between definitions}

In this section, we formalize the relation between PPKDP and APKDP and show
basic results on those definitions.
\xspace First, APKDP and PPKDP satisfy both \emph{privacy axioms} proposed
in~\cite{kifer2010towards}. These axioms express natural properties that we
expect to be true for any reasonable definition of privacy%
.

\begin{prop}\label{prop:axioms}
  PPKDP satisfies the \emph{post-processing} axiom: if a mechanism $\mecha$ is
  $\Thepsdel$-PPKDP, then for any function $f$, the mechanism $\mecha'$ defined
  by $\mecha'(D)=f\left(\mech{D}\right)$ is also $\Thepsdel$-PPKDP\@. It also
  satisfies the \emph{convexity} axiom: if two mechanisms $\mecha_1$ and
  $\mecha_2$ are both $\Thepsdel$-PPKDP, then the mechanism $\mecha$ that
  applies $\mecha_1$ with some probability $p$ and $\mecha_2$ with probability
  $1-p$ is also $\Thepsdel$-PPKDP\@.

  APKDP also satisfies these axioms.

  \iffullversion%
  \begin{proof}\label{proof:axioms}
        For APKDP, the reformulation of the definition using classical
    $\epsdel$-indistinguishability makes the result straightforward: the proof
    is the same as for $\epsdel$-DP\@. For PPKDP, we first reformulate the
    definition using $f$-divergence. Let $f(x)=\max(0,1-e^\eps x)$. Then a
    mechanism $\mecha$ is $\Thepsdel$-PPKDP iff:
    \begin{align*}
      \sum_{\hatB} \probac{\hatB}{D(i)=a} D_f\left(O_{a,\hatB}\|O_{b,\hatB}\right) \le \del
    \end{align*}
    where $O_{x,\hatB}=\mech{D}\cond{\theta,D(i)=x,\hatB}$. This view allows us
    to use the monotonicity and joint convexity properties of the $f$-divergence
    to immediately prove the result for PPKDP\@.

  \end{proof}
  \fi%
\end{prop}

We saw that $\epsdel$-APKDP bounds the probability mass of the PLRV above
$\eps$ by $\del$, for all possible partial knowledge $\hatB$. By contrast,
PPKDP bounds the same probability mass, \emph{averaged} over all possible
values of $\hatB$, weighted by their likelihood. We formalize this
interpretation and use it to show that APKDP is, as expected, stronger than
PPKDP\@. More surprisingly, we also use it to show that when $\del=0$, both
definitions are equivalent.

\begin{theorem}\label{thm:apk-ppk}
  Given a distribution $\theta$, a mechanism $\mecha$, an index $i$, two values
  $a,b\in\tuples$, an output $O$, a possible value of the partial knowledge
  $\hatB$, and a fixed $\eps>0$, let us denote
  $m\left(O,\hatB\right)=\max\left(0,1-e^{\eps-\privlossMit{a}{b}(O,\hatB)}\right)$.
  The respective quantities bounded by the requirements of APKDP and PPKDP are:
  \begin{align*}
    \APK_{i,a,b,\hatB} & = \expectu{\theta\cond{D(i)=a,\hatB},O\sim\mech{D}}{\mohb}
  \end{align*}
  and:
  \begin{align*}
    \PPK_{i,a,b} & = \expectu{\theta\cond{D(i)=a},O\sim\mech{D}}{\mob} \\
    & = \expectu{\theta\cond{D(i)=a}}{\APK_{i,a,b,B}}.
  \end{align*}
  As an immediate consequence, if $\mecha$ is $\Thepsdel$-APKDP, then it is
  also $\Thepsdel$-PPKDP\@. Further, $\Theps$-PPKDP and $\Theps$-APKDP are
  equivalent.

  \iffullversion%
    \begin{proof}
          We decompose $\PPK_{i,a,b}$ depending on $B$.
    \iffullversion%
    \begin{align*}
        \PPK_{i,a,b} 
        & = \sum_{\hatD,\hatB}\probac{\dhd,\hatB}{D(i)=a}
              \expectu{O\sim\mech{\hatD}}{\mohb} \\
        & = \sum_{\hatB}\probac{\hatB}{D(i)=a}
              \sum_{\hatD}\probac{\dhd}{D(i)=a,\hatB}
              \cdot \expectu{O\sim\mech{\hatD}}{\mohb} \\
        & = \sum_{\hatB}\probac{\hatB}{D(i)=a}
              \expectu{\theta\cond{D(i)=a,\hatB},O\sim\mech{D}}{\mohb} \\
        & = \sum_{\hatB}\probac{\hatB}{D(i)=a}\APK_{i,a,b,\hatB} \\
        & = \expectu{\theta\cond{D(i)=a}}{APK_{i,a,b,B}}.
      \end{align*}
    \else%
      \begin{align*}
        & \PPK_{i,a,b} \\
        & \; = \sum_{\hatD,\hatB}\probac{\dhd,\hatB}{D(i)=a}\expectu{O\sim\mech{\hatD}}{\mohb} \\
        & \; = \sum_{\hatB}\probac{\hatB}{D(i)=a}\sum_{\hatD}\probac{\dhd}{D(i)=a,\hatB}\\
          & \hhs \hhs \cdot \expectu{O\sim\mech{\hatD}}{\mohb} \\
        & \; = \sum_{\hatB}\probac{\hatB}{D(i)=a}\expectu{\theta\cond{D(i)=a,\hatB},O\sim\mech{D}}{\mohb} \\
        & \; = \sum_{\hatB}\probac{\hatB}{D(i)=a}\APK_{i,a,b,\hatB} \\
        & \; = \expectu{\theta\cond{D(i)=a}}{APK_{i,a,b,B}}.
      \end{align*}
    \fi{}%

    For the second part of the statement, if the mechanism is $\epsdel$-APKDP,
    then for all $i$, $a$, $b$, and $\hatB$, $\APK_{i,a,b,\hatB}\le\del$, so:
    \[
      \PPK_{i,a,b}
        = \expectu{\theta\cond{D(i)=a}}{\APK_{i,a,b,B}} 
        \le \expectu{\theta\cond{D(i)=a}}{\del}\\
        \le \del.
    \]

    For the last part of the statement, assume that $\mecha$ is
    $\Theps$-PPKDP\@. Then:
    \[
      0 = \PPK_{i,a,b} = \expectu{\theta\cond{D(i)=a}}{\APK_{i,a,b,B}}.
    \]
    All summands are non-negative, so the sum can only be $0$ if all summands
    are $0$: for all $\hatB$, $\APK_{i,a,b,\hatB}=0$, and $\mecha$ is
    $\Theps$-APKDP\@.

    \end{proof}
  \fi
\end{theorem}

When $\del>0$, the implication is strict: the PLRV can be arbitrarily higher for
certain values $\hatB$ of the background knowledge. Thus, quantifying over all
possible values $\hatB$ can lead to much larger values of $\eps$ and $\del$ than
averaging over all possible values of the background knowledge:
Example~\ref{ex:thresholding} illustrates this phenomenon. When $\del=0$
however, $\eps$-APKDP and $\eps$-PPKDP are both \emph{worst-case} properties,
like $\eps$-DP\@: an attacker's ability to choose the background knowledge does
not matter, since even for the passive attacker, we need to consider the worst
possible output $O$ and background knowledge $B$.

\subsection{$\Theta$-reducible mechanisms}

For some mechanisms and probability distributions, active attackers are, perhaps
surprisingly, not more powerful than passive attackers, even when $\del>0$. We
introduce here a necessary condition for APKDP and PPKDP to be equivalent and
we show that this condition appears in natural contexts.

Consider the example of a referendum where 2000 users take part in a vote with
two options. Each user $i$ votes ``yes'' with probability $p_i$, and ``no'' with
probability $1-p_i$. The mechanism $\mecha$ returns the exact tally of the vote.
We assume that the attacker knows half of the votes: their partial knowledge is
the vote of 1000 users. They might know that e.g. 500 of these users voted
``yes'', 500 voted ``no'', and the remaining 1000 votes are unknown. The
attacker aims to get information on the vote of her target.

Does it matter in this situation whether the attacker is passive or active? If
the attacker can choose the votes of 1000 users, she can decide that each known
user will vote ``yes''. Yet, changing these votes will only modify the tally in a
predictable way: the attacker can remove these votes from the total tally.
So, it does not matter whether these known users all vote ``yes'',
``no'', or have any other behavior known to the attacker. Without dependency
relationships between users, the attacker's uncertainty solely resides in the
unknown votes; so, a passive attacker is not weaker than an active attacker.

We generalize this intuition via the concept of \emph{$\Theta$-reducibility},
which characterizes that all possible values of background knowledge are
equivalent from a privacy perspective. We show that under this condition APKDP
is equivalent to PPKDP and formalize the above example to show that it satisfies
this condition.

\begin{definition}[$\Theta$-reducibility]\label{def:theta-reducibility}
  A mechanism $\mecha$ is $\Theta$-reducible if for all indices $i$, and all
  $B_1,B_2\in\knowledges$, there is a bijective mapping
  $\phi_{B_1,B_2}^i:\outputs\rightarrow\outputs$ such that for all $a\in\tuples$
  and for all $O\in\outputs$:
  \iffullversion%
    \begin{align*}
      \probasc{\theta}{\mech{D}=O}{D(i)=a,B=B_1}
        = \probasc{\theta}{\mech{D}=\phi_{B_1,B_2}^{i}(O)}{D(i)=a,B=B_2}.
    \end{align*}
  \fi
\end{definition}

This equivalence between possible outputs under $B_1$ and $B_2$ can be
translated to an equivalence between the corresponding the PLRVs: if $\mecha$ is
$\Theta$-reducible, then
$\privlossMit{a}{b}\left(O,B_1\right)\cond{O\sim\mech{D},D(i)=a}$ has the same
global behavior as
$\privlossMit{a}{b}\left(O,B_2\right)\cond{O\sim\mech{D},D(i)=a}$.
This equivalence between PLRVs enables us to show that $\Theta$-reducibility
implies APKDP and PPKDP are the same, even when $\del>0$. So there are mechanisms
and background knowledge functions for which an active attacker is not stronger
than a passive one.

\begin{theorem}\label{thm:apk-ppk-reducible}
  Let $\Theta$ be a family of probability distributions, and let $\mecha$ be a
  $\Theta$-reducible mechanism. Then $\mecha$ is $\Thepsdel$-APKDP iff it is
  $\Thepsdel$-PPKDP\@.

  \iffullversion%
    \begin{proof}
    First, we show that for all
    \iffullversion%
      $O\in\outputs$
    \else%
      $O$
    \fi%
    and all $B_1,B_2\in\knowledges$:
    \begin{equation*}
      \privlossMit{a}{b}(O,B_1)= \privlossMit{a}{b}(\phi_{B_1,B_2}(O),B_2).
    \end{equation*}
    This statement directly follows by unfolding the definition of
    $\privlossMit{a}{b}(O,B_1)$ (Definition~\ref{def:plrv-pk}) and
    $\Theta$-reducibility (Definition~\ref{def:theta-reducibility}):
    \iffullversion%
      \begin{align*}
        \privlossMit{a}{b}(O,B_1) 
        & = \ln\frac{\probasc{\theta}{\mech{D}=O}{D(i)=a,B=B_1}}%
                       {\probasc{\theta}{\mech{D}=O}{D(i)=b,B=B_1}} \\
        & = \ln\frac{\probasc{\theta}{\mech{D}=\phi_{B_1,B_2}(O)}{D(i)=a,B=B_2}}
                       {\probasc{\theta}{\mech{D}=\phi_{B_1,B_2}(O)}{D(i)=b,B=B_2}} \\
        & = \privlossMit{a}{b}(\phi_{B_1,B_2}(O),B_2).
      \end{align*}
    \else%
      \begin{align*}
        & \privlossMit{a}{b}(O,B_1) \\
        & \; = \ln\frac{\probasc{\theta}{\mech{D}=O}{D(i)=a,B=B_1}}%
                       {\probasc{\theta}{\mech{D}=O}{D(i)=b,B=B_1}} \\
        & \; = \ln\frac{\probasc{\theta}{\mech{D}=\phi_{B_1,B_2}(O)}{D(i)=a,B=B_2}}
                       {\probasc{\theta}{\mech{D}=\phi_{B_1,B_2}(O)}{D(i)=b,B=B_2}} \\
        & \; = \privlossMit{a}{b}(\phi_{B_1,B_2}(O),B_2)
      \end{align*}
    \fi%

    We can now prove Theorem~\ref{thm:apk-ppk-reducible}. Since $\APK_{i,a,b}$
    is the expected value of $\PPK_{i,a,b,B}$ (Theorem~\ref{thm:apk-ppk}), it is
    enough to prove that for any $B_1$ and $B_2$,
    $\APK_{i,a,b,B_1}=\APK_{i,a,b,B_2}$. Recall that
    $m\left(O,B\right)=\max\left(0,1-e^{\eps-\privlossMit{a}{b}(O,B)}\right)$,
    and fix $B_1$ and $B_2$ in $\knowledges$. Abusing notations, we denote $B_1$
    (resp. $B_2$) the event ``$B=B_1$'' (resp. $B=B_2$), $\hatD$ the event
    ``$D=\hatD$'', and $a$ the event ``$D(i)=a$''. We have:
    \iffullversion%
      \begin{align*}
        \APK_{i,a,b,B_1}
        & = \sum_{\hatO} \probac{\mech{D}=\hatO}{a,B_1} m\left(\hatO,B_1\right) \\
        & = \sum_{\hatO} \probac{\mech{D}=\phi_{B_1,B_2}(\hatO)}{a,B_2}
              m\left(\phi_{B_1,B_2}(\hatO),B_2\right)
      \end{align*}
    \else
      \begin{align*}
        &\APK_{i,a,b,B_1} \\
        & \; = \sum_{\hatO} \probac{\mech{D}=\hatO}{a,B_1} m\left(\hatO,B_1\right) \\
        & \; = \sum_{\hatO} \probac{\mech{D}=\phi_{B_1,B_2}(\hatO)}{a,B_2}
                m\left(\phi_{B_1,B_2}(\hatO),B_2\right)
      \end{align*}
    \fi
    using Definition~\ref{def:theta-reducibility} and the technical result
    above. We can then reindex the sum using the bijection
    $\hatO\rightarrow\phi_{B_1,B_2}(\hatO)$, and conclude:
    \begin{align*}
      \APK_{i,a,b,B_1}
      & = \sum_{\hatO} \probac{\mech{D}=\hatO}{a,B_2} m\left(\hatO,B_2\right) \\
      & = \APK_{i,a,b,B_2}.
    \end{align*}

    \end{proof}
  \fi
\end{theorem}

$\Theta$-reducible mechanisms are fairly common, especially under the natural
assumption that the attacker knows a fixed part of the dataset. We give a few
examples.

\begin{prop}\label{prop:reducible-mechanisms}
  Let $\Theta$ be a family of distributions in which each $\theta$ generates the
  record of each user independently, and assume that the background knowledge of
  the attacker is $k$ fixed records of the database, for a given $k$. Then the
  following mechanisms are $\Theta$-reducible.
  \begin{enumerate}
    \item \emph{Counting queries}: given a predicate $P$, the algorithm's output
      is the number of records for which $P(D(i))$ is true. This a generalizes
      binary voting.
    \item \emph{Linear queries}: given a fixed family of weights $(\alpha_i)$,
      the mechanism returns $\sum_i \alpha_i D(i)$.
    \item \emph{Different types of means}: the arithmetic mean, the geometric
      mean, the harmonic mean, and the quadratic mean (assuming all $D(i)$
      are positive).
  \end{enumerate}
  \iffullversion%
    \begin{proof}
          Fix $\tit$. For counting queries, given $B_1,B_2\in\knowledges$, define
    $\phi_{B_1,B_2,i}^{count}(\mech{D})=\mech{D}-\mech{B_1}+\mech{B_2}$. This
    function is linear, so injective. Call $\theta'$ the random distribution of
    the $n-k$ records not present in $B$. Then for all $a\in\tuples$ and all
    $O\in\outputs$:
    \begin{align*}
      & \probasc{\theta}{\mech{D}=O}{D(i)=a,B=B_1} \\
      & \; = \probasc{D'\sim\theta'}{\mech{D' \cup B_1}=O}{{(D'\cup B_1)}_i=a} \\
      & \; =\probasc{D'\sim\theta'}{\mech{D' \cup B_2}=\phi_{B_1,B_2,i}^{count}(O)}{{(D'\cup B_2)}_i=a} \\
      & \; =\probasc{\theta}{\mech{D}=\phi_{B_1,B_2,i}^{count}(O)}{D(i)=a,B=B_2}.
    \end{align*}

    For linear queries, we use a similar mapping to $\phi$, which also depends
    on the mapping. Let $I$ be the indices of records present in $B$; then a
    linear query is $\Theta$-reducible with:
    \begin{align*}
      \phi_{B_1,B_2,i}^{linear}(\mech{D})
        = \mech{D}+\sum_{j\in I} \alpha_j\left({\left(B_2\right)}_j-{\left(B_1\right)}_j\right).
    \end{align*}
    This also proves the result for the arithmetic mean.

    Similarly, it is easy to verify that the following $\phi$ functions show
    $\Theta$-reducibility for the geometric, harmonic and quadratic mean.
    \begin{align*}
     & \phi_{B_1,B_2,i}^{geometric}(\mech{D})
        = {\left(\mech{D}^n\cdot\frac{\prod_{j\in I}{\left(B_2\right)}_j}{\prod_{j\in I}{\left(B_1\right)}_j}\right)}^{1/n} \\
     & \phi_{B_1,B_2,i}^{harmonic}(\mech{D}) \\
     & \hs = n{\left({\left(\frac{\mech{D}}{n}\right)}^{-1}
         + \sum_{j}\left(\frac{1}{{\left(B_2\right)}_j}-\frac{1}{{\left(B_1\right)}_j}\right)\right)}^{-1} \\
     & \phi_{B_1,B_2,i}^{quadratic}(\mech{D}) \\
     & \hs = \sqrt{n\cdot\mech{D}^2 + \sum_j\frac{{\left(B_2\right)}_j^2-{\left(B_1\right)}_j^2}{n}}
    \end{align*}
    The injectivity of each of these functions is clear.

    \end{proof}
  \fi
\end{prop}

Note that in Proposition~\ref{prop:reducible-mechanisms}, it is crucial that the
records in the attacker's partial knowledge are \emph{fixed}. In general,
knowing the records of $k$ users is \emph{not} equivalent to knowing the records
of $k$ other users. Indeed, if these records are generated with different
probabilities, the randomness from the unknown records differs between both
scenarios, and it is in general impossible to convert one into the other.

\begin{remark}
  Observation~1 in~\cite{grining2017towards} claims that the partial knowledge
  of $k$ records in a database of size $n$ is the same as no partial knowledge
  in a database of size $n-k$. For counting queries, this holds for the same
  reason that counting queries are $\Theta$-reducible: one can ``remove'' the
  $k$ known records from the mechanism output and obtain a bijection between the
  cases with and without partial knowledge. Thus, the partial knowledge is
  irrelevant to the mechanism's privacy and can be ignored.
  
  This observation, however, does not hold in general: we later show in
  Example~\ref{ex:thresholding} that counting queries \emph{with thresholding}
  are not $\Theta$-reducible, and in this case, the knowledge of $k$ records in
  a database of size $n$ has a very different effect than no partial knowledge
  in a database of size $n-k$.
\end{remark}

\section{Applications}\label{sec:applications}

In this section, we show that the foundations layed in the previous sections can
be applied to practical problems. First, we propose improved bounds on the
privacy of noiseless counting queries under partial background knowledge.
Second, we investigate counting queries with \emph{thresholding} and show that
thresholding can be used to improve privacy for these queries.

These two results formalize intuitions that privacy practitioners frequently use
when trying to assess the risk of releasing aggregated data.
Theorem~\ref{thm:binary-voting} provides a rigorous explanation of why, under
reasonable assumptions on the background knowledge of a realistic attacker,
counting queries over a large user population do not necessarily leak individual
information. Theorem~\ref{thm:thresholding} shows that thresholding provides
protection in those cases where a counting query only captures a small number of
users; this provides a new interpretation of this technique as giving formal
privacy guarantees when the attacker has partial background knowledge.

\subsection{Counting queries}\label{sec:counting}

The initial motivation for limiting the attacker's background knowledge was to
show that, under this assumption, some noiseless mechanisms preserve the
individuals' privacy~\cite{bhaskar2011noiseless}. A typical example is a
\emph{counting query}, which answers the question ``How many users satisfy
$P$?'' for some property $P$. We can model this by a data-generating
distribution $\theta$ where each record $D(i)$ is either $0$ or $1$ with some
probability $p_i$, and we want to measure the privacy of the mechanism
$\mech{D}=\sum_i D(i)$. Records are assumed to be independent, and the adversary
is assumed to know some portion of the records. As an immediate consequence of
Theorem~\ref{thm:apk-ppk-reducible} and
Proposition~\ref{prop:reducible-mechanisms}, it does not matter whether the
attacker can modify, or only see, this portion of records: the values of $\eps$
and $\del$ are identical for APKDP and PPKDP\@.

Furthermore, the closer $p_i$ are to $0$ or $1$, the less randomness is present
in the data. For extremely small or large values of $p_i$, the situation is very
similar to one where the attacker exactly knows $D(i)$. As such, it is natural
to assume that among the records that are unknown by the attacker, all $p_i$ are
between $\lambda$ and $1-\lambda$, for some $\lambda$ not too close to $0$. This
assumption can easily be communicated to non-specialists: ``we assume that there
are at least 1000 records that the attacker does not know, and that her level of
uncertainty is at least 10\% for these records.''

Initial asymptotic results in this context appeared
in~\cite{bhaskar2011noiseless} and more precise bounds were derived
in~\cite{grining2017towards}. In the special case where all $p_i$ are equal to a
fixed value $p$, Theorem~5 in~\cite{bhaskar2011noiseless} and Theorem~1
in~\cite{grining2017towards} show that counting queries are APKDP with
$\eps=O\left(\sqrt{\frac{\ln(1/\del)}{n}}\right)$ (for small $\del$, and
increasing $n$) and $\del=e^{-\Omega\left(\eps^2n\right)}$ (for small $\eps$,
and increasing $n$). This provides tiny values of $\eps$ and $\del$ for moderate
values of $n$ and $p$. However, the assumption that all $p_i$ are identical is
unrealistic: in practice, there is no reason to assume that all users have an
equal chance of satisfying $P$. Theorem~7 in~\cite{bhaskar2011noiseless} and
Theorem~2 in~\cite{grining2017towards} show that without this assumption, the
$\eps$ obtained is still small: $\eps=O\left(\sqrt{\frac{\ln(n)}{n}}\right)$,
but the upper bound obtained on $\del$ is significantly larger:
$\del=O\left(\frac{1}{\sqrt{n}}\right)$. This is more than what is typically
acceptable; a common recommendation is to choose
$\del=o\left(\frac{1}{n}\right)$.

In the following theorem, we show that the exponential decrease of $\del$ with
$n$ still holds in the general case where all $p_i$ are different. For
simplicity, we assume that the attacker has no background knowledge: because all
records are independent, adding some partial knowledge has a fixed, reversible
effect on the output space, similarly to $\Theta$-reducibility. In this case,
having the attacker know $m$ records out of $n$ is the same as having the
attacker know no records among $n-m$.

\begin{theorem}\label{thm:binary-voting}
  Let $\theta$ be a distribution that generates $n$ records, where $D(i)$ is the
  result of an independent Bernoulli trial of probability $p_i$. Let $\lambda$
  be such that for all $i$, $\lambda<p_i<1-\lambda$. Let $\mecha$ be defined by
  $\mech{D}=\sum_i D(i)$. Then $\mecha$ is $\Thepsdel$-APKDP, for any $\eps$ and
  $\del$ such that:
  \begin{equation*}
    \del\ge\proba{\frac{X}{Y}\ge\eps}
  \end{equation*}
  where $X$ and $Y$ are independent random variables sampled from a binomial
  distribution with $n-1$ trials and success probability $2\lambda$. For a fixed
  $\eps\le1$, this condition is satisfied if:
  \begin{equation*}
    \eps \ge \max\left(
                \sqrt{\frac{14\ln(1/\del)}{\lambda(n-1)}},
                \frac{27}{\lambda(n-1)}\right)
  %
  \end{equation*}
  \iffullversion%
  which gives $\del=e^{-\Omega\left(\eps^2\lambda n\right)}$.
  \fi%
  \iffullversion%
    \begin{proof}
The proof uses existing results on privacy amplification by shuffling:
in~\cite{erlingsson2019amplification,balle2019privacy}, the authors show
that adding noise independently to each data point, and then shuffling the
results (hiding from the attacker which record comes from which user),
provide strong DP guarantees. Even though our problem looks different, the
same reasoning can be applied. First, we show that $\theta$ can be seen as
applying randomized response on each record. Then, since a counting query is
a symmetric boolean function, it can be composed with a shuffle of its
input, which allows us to use amplification by shuffling.

Let us formalize this intuition. A Bernoulli trial of probability $p_i$
(denoted $\ber{p_i}$), with $\lambda<p_i<1-\lambda$, can be decomposed into
the following process:
\begin{itemize}
  \item Generate $b\sim\ber{2\lambda}$.
  \item If $b=0$, return $b_{\nu_i}\sim\ber{\frac{p_i-\lambda}{1-2\lambda}}$.
  \item If $b=1$, return $b_{rr}\sim\ber{0.5}$.
\end{itemize}
This can be seen as a \emph{randomized response} process applied on some
input $b_{\nu_i}$, itself random: $\theta=\rrdl(\nu)$, where $\nu$ is a
distribution generating $n$ records, where the $i$-th record is generated by
$b_{\nu_i}\sim\ber{\frac{p_i-\lambda}{1-2\lambda}}$, and $\rrdl$ is a binary
randomized response process with parameter $2\lambda$.

Note that $\mecha$ can be seen as the composition between itself
and a pre-shuffling phase: $\mecha=\mecha\circ\shuffle$, where
$\shuffle:\datasets\rightarrow\datasets$ is a function that applies a random
permutation to the input records. Thus,
$\mech{D}\cond{D\sim\theta}=\mech{\shuffl{\rrdl(D)}}\cond{D\sim\nu}$. We can
now apply Theorem~3.1 in~\cite{balle2019privacy} and its proof to show that
$\shuffle\circ\rrdl$ is $\epsdel$-DP for $\eps$ within the constraints above
(with $k=2$ and $\gamma=2\lambda$). By post-processing,
$\mecha\circ\shuffle\circ\rrdl$ is also $\epsdel$-DP, which directly yields
that $\mecha$ is $\epsdel$-APKDP\@.

Note that we omitted a small technical detail: conditioning $\theta$ on
$D(i)=a$ is not identical to conditioning $\nu$ on $D(i)=a$, since no noise
is added to the record $i$ in the former case. To fix this, we need to
define $\rrdl$ as randomizing all records except a fixed one $i$. The proof
of Theorem~3.1 in~\cite{balle2019privacy} assumes that no noise is added to
the target record, so the result still holds.

    \end{proof}
  \fi
\end{theorem}


We compare this result with the previous state-of-the-art. First, we reformulate
a previously known result from~\cite{grining2017towards} that applies to our
setting.

\begin{prop}[Theorem~3 in~\cite{grining2017towards}]\label{prop:binary-voting}
  Let $\theta$ be a distribution that generates $n$ records, where $D(i)$ is the
  result of an independent Bernoulli trial of probability $p_i$, and let $\mecha$
  be defined by $\mech{D}=\sum_i D(i)$. Let
  $\mu_2=\frac{1}{n}\sum_i p_i\left(1-p_i\right)$ and
  $\mu_3=\frac{1}{n}\sum_i p_i\left(1-p_i\right)\left|1-2p_i\right|$ be
  respectively the average second moment and average absolute third moment of the
  $D(i)$.
  Then for any $\del_2\ge1.25e^{-n\mu_2/2}$, $\mecha$ is
  $\thepsdel$-APKDP, with
  \begin{equation*}
    \eps=\sqrt{\frac{2\ln(1.25/\del_2)}{n\mu_2}}
    \quad\text{ and }\quad
    \del=\frac{1.12\mu_3}{\sqrt{n}\mu_2^3}\left(1+e^\eps\right) + \del_2.
  \end{equation*}

  \begin{proof}
    For $\del_2=\frac{4}{5\sqrt{n}}$, this is a direct application of Theorem~3
    in~\cite{grining2017towards}. Changing the value $\del_2$ in its proof
    (Appendix~A.2) allows us to obtain the more general formula above. This
    requires Fact~1 to be true, which is the case when $\eps\le1$ (the authors
    omit this detail), or equivalently, when $\del_2\ge1.25e^{-n\mu_2/2}$.
  \end{proof}
\end{prop}

\begin{figure*} 
  \begin{center}
  \begin{tikzpicture}[scale=\figurescale]
    \begin{axis}[
        xmin=0,
        xmax=2,
        ymode=log,
        ymin=1e-10,
        ymax=2,
        xlabel=$\eps$,
        ylabel=$\del$,
        no markers,
        legend pos=south west,
        legend cell align={left}
      ]
      \addplot[
        blue,
        dashed
      ] table [x=epsilon, y=delta, col sep=comma]{comparison_n=100_good.csv};
      \addplot[
        red,
        dashed
      ] table [x=epsilon, y=delta, col sep=comma]{comparison_n=100_realist.csv};
      \addplot[
        red,
        solid
      ] table [x=epsilon, y=delta, col sep=comma]{comparison_n=100_closed.csv};
      \addplot[
        blue,
        solid
      ] table [x=epsilon, y=delta, col sep=comma]{comparison_n=100_numeric.csv};
      \legend{
        \autoref{prop:binary-voting}, case 1 \\
        \autoref{prop:binary-voting}, case 2 \\
        \autoref{thm:binary-voting}, closed-form \\
        \autoref{thm:binary-voting}, numeric \\
      }
    \end{axis}
  \end{tikzpicture}
  \begin{tikzpicture}[scale=\figurescale]
    \begin{axis}[
        xmin=0,
        xmax=1,
        ymode=log,
        ymin=1e-15,
        ymax=2,
        xlabel=$\eps$,
        ylabel=$\del$,
        no markers,
      ]
      \addplot[
        blue,
        dashed
      ] table [x=epsilon, y=delta, col sep=comma]{comparison_n=10000_good.csv};
      \addplot[
        red,
        dashed
      ] table [x=epsilon, y=delta, col sep=comma]{comparison_n=10000_realist.csv};
      \addplot[
        red,
        solid
      ] table [x=epsilon, y=delta, col sep=comma]{comparison_n=10000_closed.csv};
      \addplot[
        blue,
        solid
      ] table [x=epsilon, y=delta, col sep=comma]{comparison_n=10000_numeric.csv};
    \end{axis}
  \end{tikzpicture}
  \vspace{-1em}
  \end{center}
  \caption{Comparison of $\epsdel$ bounds given by
    Theorem~\ref{thm:binary-voting} and Proposition~\ref{prop:binary-voting},
    for $\lambda=0.05$, $n=\num{100}$ (left) and $n=\num{10000}$ (right).\\
    Case 1: all but one $p_i$ are 0.5. Case 2: the $p_i$ are distributed
    uniformly over $[0.05,0.95]$.}\label{fig:comparison}
\end{figure*}
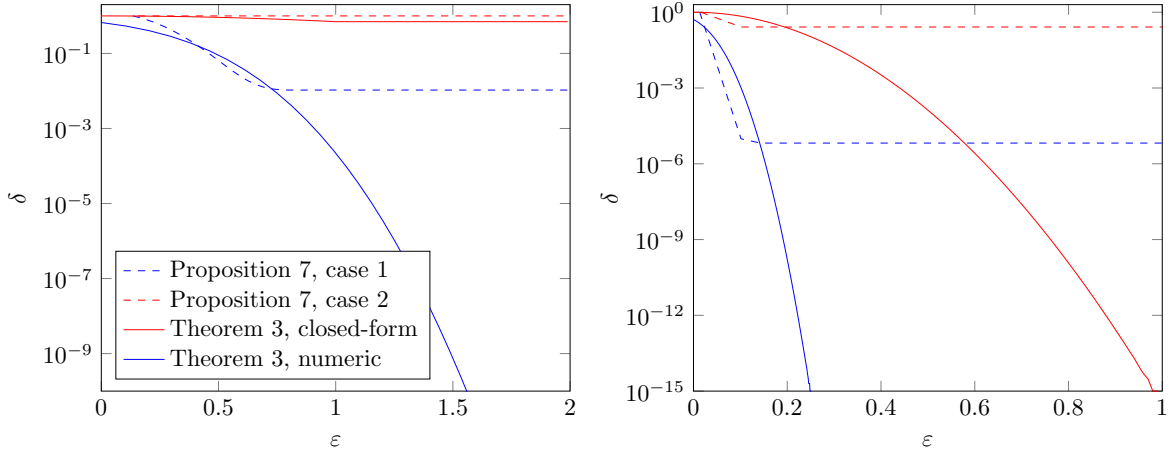

The comparison between this result and Theorem~\ref{thm:binary-voting} is not
completely straightforward. Aside from $n$, Theorem~\ref{thm:binary-voting} only
depends on a global bound on the ``amount of randomness'' ($p_i$) of each user,
while Proposition~\ref{prop:binary-voting} depends on the average behavior of
all users. As such, the global bound $\lambda$ can be small because of one
single user having a low $p_i$, even if all other users have a lot of variance
because their $p_i$ is close to 0.5. We therefore provide two experimental
comparisons. In the first one, $p_1=\lambda=0.05$ and $p_i=0.5$ for all $i>1$.
This case is designed to have the parameters of Theorem~\ref{thm:binary-voting}
underperform (as we underestimate the total amount of randomness) and those of
Proposition~\ref{prop:binary-voting} perform well. In the second one, the $p_i$
are uniformly distributed in
$\left[\lambda,1-\lambda\right]=\left[0.05,1-0.05\right]$. In both cases, we
compare the $\epsdel$ graphs obtained for $n=10^3$ and $n=10^5$, and present the
results in Figure~\ref{fig:comparison}.

The graphs show that if we consider the smallest possible $\eps$ given by the
definitions, our theorem leads to a large $\del$: with
$\eps=\Theta\left(1/\sqrt{n}\right)$, we obtain $\del=\Theta(1)$; in contrast,
Proposition~\ref{prop:binary-voting} leads to
$\del=\Theta\left(1/\sqrt{n}\right)$. However, increasing $\eps$ to slightly
larger values quickly leads to tiny values of $\del$, which was impossible with
the previous state-of-the-art results. They also show that the closed-form bound
from~\cite{balle2019privacy} is far from tight, as numerically computating these
bounds improves them by several orders of magnitude. This leads to a natural
open question: is there a better asymptotic formulation of the bounds given by
amplification by shuffling for randomized response?

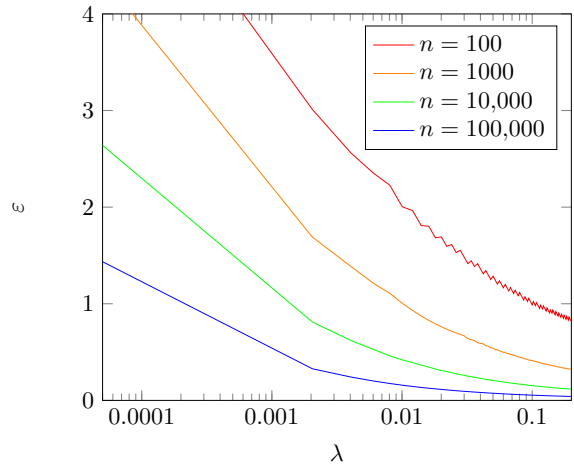
\begin{figure}[t] 
  \begin{center}
  \begin{tikzpicture}[scale=\figurescale]
    \begin{axis}[
        xmin=0.00005,
        xmax=0.2,
        xmode=log,
        ymin=0,
        ymax=4,
        log ticks with fixed point,
        xlabel=$\lambda$,
        ylabel={$\eps$},
        no markers,
        legend pos=north east,
        legend cell align={left}
      ]
      \addplot[
        red,
        solid
      ] table [x=lambda, y=epsilon, col sep=comma]{comparison-p_n=100.csv};
      \addplot[
        orange,
        solid
      ] table [x=lambda, y=epsilon, col sep=comma]{comparison-p_n=1000.csv};
      \addplot[
        green,
        solid
      ] table [x=lambda, y=epsilon, col sep=comma]{comparison-p_n=10000.csv};
      \addplot[
        blue,
        solid
      ] table [x=lambda, y=epsilon, col sep=comma]{comparison-p_n=100000.csv};
      \legend{$n=100$, $n=\num{1000}$, $n=\num{10000}$, $n=\num{100000}$}
    \end{axis}
  \end{tikzpicture}
  \vspace{-1em}
  \end{center}
  \caption{Comparison of $\eps$ bounds given by the numerical computation
    of \autoref{thm:binary-voting}, with varying $\lambda$, for various values
    of $n$ and $\del=0.01/n$.}\label{fig:varying-lambda}
\end{figure}

What is the impact of $\lambda$ on the privacy guarantees? In
\autoref{fig:varying-lambda}, we plot the $\eps$ obtained for $\del=0.01/n$ as a
function of $\lambda$, for various values of $n$.

One natural application for this result is \emph{voting}: in typical elections,
the total tally is released without any noise. Adding noise to the election
results, or not releasing them, would both be unacceptable. Thus, the results
are not $\epsdel$-DP for any $\epsdel$ parameters, even though publishing the
tally is not perceived as a breach of privacy. The intuitive explanation for
this is that attackers are assumed not to have complete background knowledge of
the secret votes. Our results confirm this intuition and quantify it. These
results can easily be extended to votes between multiple candidates.

\begin{corollary}\label{cor:generalized-voting}
  Let $\theta$ be a distribution that generates $n$ records, where
  $D(i)\in\{1,\dots,K\}$ for $K>1$, and where $\proba{D_i=k}=p_{i,k}$, and every
  record is independent from all others. Let $\lambda$ be such that for all $i$
  and all $k$, $\lambda<p_{i,k}$. Let $\mecha$ return the histogram of all
  values: $\mech{D}=\left(N_1,\dots,N_K\right)$, where $N_k$ is the number of
  records $i$ such that $D(i)=k$. Then $\mecha$ is $\Thepsdel$-APKDP, for any
  $\eps\le1$ and $\del>0$ such that:
  \begin{equation*}
    \eps \ge \max\left(
                \sqrt{\frac{14\ln(1/\del)}{\lambda(n-1)}},
                \frac{27}{\lambda(n-1)}\right).
  \end{equation*}
\end{corollary}

  \begin{proof}
    The proof is the same as for Theorem~\ref{thm:binary-voting}. With multiple
    options, the parameter $\gamma$ of the multi-category randomized response is
    $\gamma=K\lambda$, which leads to the same $\epsdel$ parameters.
  \end{proof}

The results in this section apply to \emph{individual} counting queries. This
applies to scenarios like votes, but in many practical use cases, multiple
queries are released. Can the results of this section be generalized to these
cases? In general, noiseless mechanism do not compose. For example, fixing an
individual $t$, queries like ``How many people voted 1?'' and ``How many people
who are not $t$ voted 1?'' can both be private on their own. However, publishing
both results will reveat $t$'s vote: the composition of both queries cannot be
private. Are there special cases where noiseless counting queries can be
composed?

One such case happens when each counting query contains the data of a number of
\emph{new} users, independent from users in previous counting queries. This can
happen in situations where statistics are collected on actions that each user
can only do once, for example, registering with an online service. In this case,
we can restrict the privacy analysis of each new query to the set of independent
users in its input, and use the previous results from this section: this
approach is formalized and proven in Theorem~1 in~\cite{bhaskar2011noiseless}.

What if this approach is impossible, for example if there are dependencies
between each input record of each query? For example, a referendum could ask
voters \emph{multiple} questions, with correlations between the different
possible answers. Another example could be app usage statistics published every
day, where the data for day $d+1$ for each user is correlated to the user's data
on the previous day $d$. In this case, to compute the privacy loss of the first
$d$ binary queries, we can consider them as a \emph{single} query with $2^d$
options. Afterwards, we can take into account the temporal correlations to
compute the probabilities associated with each option, and use
Corollary~\ref{cor:generalized-voting}.

\subsection{Thresholding}\label{sec:thresholding}

Theorem~\ref{thm:binary-voting} gives good $\eps$ and $\del$ parameters when
there are many people who vote with ``enough randomness'': there is a $\lambda$
such that $\lambda<p_i<1-\lambda$. The parameters have a dependency on $\lambda
n$, which in practice translates to scenarios where both options have large
counts with high probability. In many practical applications, however, it is
hard to know in advance whether this will be the case. Consider, for example, a
mobile app gathering usage metrics on possible sequences of actions carried by
users within the app. Some sequences will be very probable, and have high
counts. But if there are arbitrarily many such sequences, some will be very
rare: like many practical distributions, there will be a \emph{long tail}.

To protect the data from these outlier users, a typical protection employed is
\emph{thresholding}: only return the user count associated with a sequence if it
is larger than a given threshold $T$. What level of protection does such a
technique provide? In this section, we formalize the intuition given in
Example~\ref{ex:thresholding}, and show that, assuming a passive attacker,
thresholding provides protection when all voters vote with a very small or a
very large probability. First, we formalize the notion of a thresholding
mechanism in a simple context.

\begin{definition}[Simple thresholding]
  Given a database $D=(D(1),\ldots,D(n))$ with values in $\{0,1\}$ and a
  threshold $T$, the \emph{$T$-thresholding mechanism} $\mecha_T$ evaluates
  $\tilde{k}=\sum_i D(i)$ and returns $\bot$ if $\tilde{k} \le T$, and $k$
  otherwise.
\end{definition}

Note that $\mecha_T$ only thresholds \emph{low} counts. In many practical
situations, however, thresholding is applied in both directions: it must
also catch the case where, with high probability, almost all records are $1$.
This situation is symmetrical to thresholding low counts: without loss of
generality, we can assume $T<n/2$, and the symmetric version of all results in
this section hold.

Let us now show the main result of this section: if participants vote $1$ with
low probability, then thresholding protects against passive attackers, and in
some cases also against certain active attackers. This privacy property only
holds if the expected value of the count is lower than the threshold; and the
level of protection depends on the ratio between the threshold and the expected
value (denoted by $r$ below).

\begin{theorem}\label{thm:thresholding}
  Let $\theta$ be a distribution that returns $n$ independent records, each of
  which is $1$ with low probability: $D(i)\sim Ber(p_i)$, and $p_i<p$ for all
  $i$; moreover, let $\Theta=\{\theta\}$. Suppose that there is no partial
  knowledge, i.e., $|B|=0$, and let us denote by $f(s,n,p)$ the probability that
  a random variable following a binomial distribution with parameters $n$ and
  $p$ has value $s$: $f(s,n,p)=p^s{(1-p)}^{n-s}\binom{n}{s}$.

  Then, if $r=\frac{p(n-1)}{(1-p)T}<1$, $\mecha_T$ is $\Thepsdel$-APKDP (and
  thus, $\Thepsdel$-PPKDP), with:
  \begin{align*}
    \eps & = -\ln\left(1-\frac{f(T,n-1,p)}{1-r}\right) & \del = \frac{f(T,n-1,p)}{1-r}
  \end{align*}
  For a large $n$, assuming $pn$ is fixed, we can use the Poisson approximation
  and get $\del\approx\frac{{(pn)}^T e^{-pn}}{(1-r)T!}$. If this quantity is
  small enough, $\eps\approx\del$.

  If the background knowledge $B$ is not empty, assume that the attacker knows a
  subset $|B|$ of records. Let $b_{\max}$ be such that
  $r_b=\frac{p|B|}{(1-p)b_{\max}}<1$ and
  $r'=\frac{p(n-|B|-1)}{(1-p)(T-b_{\max})}<1$. Then $\mecha_T$ is
  $\Thepsdel$-PPKDP, with:
    \begin{align*}
      \eps & = -\ln\left(1-\frac{f(T-b_{\max},n-|B|-1,p)}{1-r'} \right) \\
      \del & = \frac{f(b_{\max},|B|,p)}{1-r_b} + \frac{f(T-b_{\max},n-|B|-1,p)}{1-r'}.
    \end{align*}

  \begin{proof}
    The proof is presented in three stages.
    \begin{enumerate}
		\item First, we consider the simpler case where all $p_i$ are equal and
        there is no background knowledge. This allows us to compute the PLRV
        exactly, and we can then split the output space into two parts. Most of
        its mass will be in the $\bot$ event, and we can compute it there
        exactly. All other events are captured by $\del$.
		\item Second, we extending this to non-empty partial knowledge in a
        similar fashion: for some $b_{\max}$, with high probability, the
        background knowledge will not have more than $b_{\max}$ records whose
        value is 1: the rest of the probability mass goes in the $\del$, and this
        allows us to use the previous idea with a new threshold $T'=T-b_{\max}$.
		\item Finally, we use a coupling argument to extend this to the case where
      the $p_i$ are not all the same.
    \end{enumerate}
    \iffullversion%
          First, let us compute the PLRV for $\mecha_T$ depending on the output $k$
    and the value of the background knowledge $\hatB$, assuming a simple
    distribution $\theta$ where records are i.i.d\@. Denote by $b$ the number of
    records in $\hatB$ which are 1 and by $\ol{b}=|\hatB|-b$ the number of
    records that are 0. The targeted record will be called $D(t)$, and we assume
    it is never part of $\hatB$. Let $\theta$ be a distribution that returns $n$
    i.i.d records according to $D(i)\sim Ber(p)$. Then we can directly compute:
    \begin{align*}
      \privlossit{0}{1}{\mecha_T}\left(k,\hatB\right)
      & = \ln \frac{\probac{\mech{D}=k}{D(t)=0,B=\hatB}}
                   {\probac{\mech{D}=k}{D(t)=1,B=\hatB}} \\
      & = \begin{cases*}
            \ln\frac{\sum_{s=0}^{T-b} f(s, n-|B|-1, p)}
                    {\sum_{s=0}^{T-b-1} f(s, n-|B|-1, p)}
              & if $k=\bot$ \\
            \ln\frac{p(n-\ol{b}-k)}{(1-p)(k-b)}
              & otherwise.
          \end{cases*}
    \end{align*}
    Note that if $k=b$, then $\privlossit{0}{1}{\mecha_T}(k,B)=\infty$. The case
    where $k<b$ is impossible regardless of $D(i)$: there cannot be more $1$s in
    the background knowledge than the mechanism outputs. In the case where there
    is no background knowledge, this becomes:
    \begin{align*}
      \privlossit{0}{1}{\mecha_T}(k)
      = \begin{cases*}
          \ln\frac{\sum_{s=0}^{T} f(s, n-1, p)}
                  {\sum_{s=0}^{T-1} f(s, n-1, p)}
            & if $k=\bot$ \\
          \ln\frac{p(n-k)}{(1-p)k}
            & otherwise.
        \end{cases*}
    \end{align*}
    This calculation allows us to bound $\eps$ and $\del$ when there is no
    background knowledge. First, we need a technical lemma to bound the
    probability mass of the tail of the binomial distribution appearing above.

    \begin{lemma}\label{lem:partial-binomial}
      For any $n$, $p$, and $m$ such that $m>\frac{pn}{1-p}$:
      \begin{align*}
        \sum_{s=m}^n f(s,n,p) < \frac{f(m,n,p)}{1-\frac{pn}{(1-p)m}}.
      \end{align*}
      \begin{proof}
        Note that for all $s \ge m$:
        \begin{align*}
          \frac{f(s+1,n,p)}{f(s,n,p)} & = \frac{p}{1-p}\left(\frac{n-s}{s+1}\right) < \frac{pn}{(1-p)m}.
        \end{align*}
        Since $m>\frac{pn}{1-p}$, this is strictly lower than $1$, so the sum
        converges at least as fast as a geometric series, which directly gives
        the desired result.
      \end{proof}
    \end{lemma}

    Now, let us prove the main theorem when there is no background knowledge and
    all $p_i$ are equal. We need to consider both $\privlossit{1}{0}{\mecha_T}$
    and $\privlossit{0}{1}{\mecha_T}$. First, we have:
    \begin{align*}
      \probasc{\theta}{\mecha_T(D)\neq\bot}{D(i)=0}
      & < \probasc{\theta}{\mecha_T(D)\neq\bot}{D(i)=1} \\
      & = 1 - \sum_{s=0}^{T-1} f(s,n-1,p) \\
      & = \sum_{s=T}^{n-1} f(s,n-1,p) \\
      & < \frac{f(T,n-1,p)}{1-\frac{p(n-1)}{(1-p)T}}
    \end{align*}
    since $T>\frac{p(n-1)}{1-p}$, so we can use
    Lemma~\ref{lem:partial-binomial}. Let us denote this quantity as $\del$.
    Now, we have
    \begin{align*}
    \privlossit{1}{0}{\mecha_T}(\bot)
      = \ln\frac
          {\sum_{s=0}^{T-1} f(s,n-1,p)}
          {\sum_{s=0}^{T} f(s,n-1,p)}
      < 0.
    \end{align*}
    Furthermore:
    \begin{align*}
      \privlossit{0}{1}{\mecha_T}(\bot)
      & = \ln\frac
            {\sum_{s=0}^{T} f(s,n-1,p)}
            {\sum_{s=0}^{T-1} f(s,n-1,p)} \\
      & < \ln\left(\frac
            {1}
            {1-\sum_{s=T}^{n-1} f(s,n-1,p)}\right) \\
      & < -\ln\left(1-\frac{f(T,n-1,p)}{1-\frac{p(n-1)}{(1-p)T}}\right).
    \end{align*}
    Thus, when the output is thresholded, the PLRV is smaller than
    $\eps=-\ln\left(1-\frac{f(T,n-1,p)}{1-r}\right)$, and the event ``the output
    is not thresholded'' only happens with a probability smaller than $\del$,
    which proves the initial statement in the simpler case.

    Now, in the case where the background knowledge is non-empty, we must not
    only split the output space, but also $\mathcal{B}$ as well. Denoting $b$
    the number of ``1'' entries in $B$, there are three cases we must consider:
    \begin{enumerate}[nosep]
      \item $b \ge b_{\max}$: if $b_{\max}$ is large enough, this happens with
        small probability, which we put in the $\del$ term;
      \item $b<b_{\max}$ and $\mecha_T(D)\neq\bot$: if $T'=T-b_{\max}$ is large
        enough, this happens with small probability, which we put in the $\del$;
      \item $b<b_{\max}$ and $\mecha_T(D)=\bot$: this is the event in which most
        of the probability mass is concentrated on, so we bound its privacy loss
        to obtain $\eps$.
    \end{enumerate}
    The probability of the first event can be bounded by:
    \begin{align*}
      \probas{\theta}{b \ge b_{\max}}
      & = \sum_{s=b_{\max}}^{|B|} f(s,|B|,p) \\
      & < \frac{f(b_{\max},|B|,p)}{1-\frac{p|B|}{(1-p)b_{\max}}}
    \end{align*}
    by Lemma~\ref{lem:partial-binomial}. Similarly, the probability of the
    second event can be bounded by:
    \iffullversion%
      \begin{align*}
        \probasc{\theta}{\mecha_T(D)\neq\bot}{b < b_{\max}, D(i)=1}
        & < \sum_{s=T'}^{n-|B|-1} f(s,n-|B|-1,p) \\
        & < \frac{f(T',n-|B|-1,p)}{1-\frac{p(n-|B|-1)}{(1-p)T'}}
      \end{align*}
    \else%
      \begin{align*}
        & \probasc{\theta}{\mecha_T(D)\neq\bot}{b < b_{\max}, D(i)=1} \\
        & \quad < \sum_{s=T'}^{n-|B|-1} f(s,n-|B|-1,p) \\
        & \quad < \frac{f(T',n-|B|-1,p)}{1-\frac{p(n-|B|-1)}{(1-p)T'}}
      \end{align*}
    \fi%
    so we can bound $\del$ by the sum of those two terms. Now, let us compute
    the privacy loss for the third case. Assuming $b<b_{\max}$, we have:
    \begin{align*}
      \privlossit{1}{0}{\mecha_T}(\bot,\hatB)
        = \ln\frac
            {\sum_{s=0}^{T-b-1} f(s,n-|B|-1,p)}
            {\sum_{s=0}^{T-b} f(s,n-|B|-1,p)}
        < 0
    \end{align*}
    and:
    \begin{align*}
      \privlossit{0}{1}{\mecha_T}(\bot,\hatB)
      & = \ln\frac
            {\sum_{s=0}^{T-b} f(s,n-|B|-1,p)}
            {\sum_{s=0}^{T-b-1} f(s,n-|B|-1,p)} \\
      & < \ln\left(\frac
            {1}
            {\sum_{s=0}^{T'-1} f(s,n-|B|-1,p)}\right) \\
      & < -\ln\left(1 - \frac{f(T',n-|B|-1,p)}{1-r'}\right).
    \end{align*}
    Denoting this by $\eps$, this proves the theorem in the special case where
    all $p_i$ are equal to a constant $p$.

    Now, we extend the first case (where the background knowledge is empty) to
    the case where all $p_i$ are different, and $p_i<p$ for all $p$. Let us
    denote $\theta_p$ the distribution where all users vote with the same
    probability $p$. Let $g(s,n,p)=\probas{\theta}{\mecha_T(D)=s}$. By a simple
    coupling argument between $\theta$ and $\theta_p$, we have for all $t$:
    \begin{align*}
      \sum_{s=t+1}^n g(s,n,p)
      & = \probas{\theta}{\mecha_T(D)>t} \\
      & \le \probas{\theta_p}{\mecha_T(D)>t} \\
      & = \sum_{s=t+1}^n f(s,n,p).
    \end{align*}
    We can then use this fact throughout the previous proof. This is immediate
    for $\del$, and for $\eps$, we have:
    \begin{align*}
      \privlossit{0}{1}{\mecha_T}(\bot,\hatB)
      = \ln\frac
          {\sum_{s=0}^{T} g(s,n-1,p)}
          {\sum_{s=0}^{T-1} g(s,n-1,p)}.
    \end{align*}
    Bound the numerator by $1$ and expand the denominator:
    \begin{align*}
      \sum_{s=0}^{T-1} g(s,n-1,p)
      & = 1 - \sum_{s=T}^{n-1} g(s,n-1,p) \\
      & \ge 1 - \sum_{s=T}^{n-1} f(s,n-1,p) \\
      & = \sum_{s=0}^{T-1} f(s,n-1,p)
    \end{align*}
    so we can reuse the previous bound. The bounds translate to the case where the
    background knowledge is not empty by a similar argument.
    \fi%
  \end{proof}
\end{theorem}

\begin{figure}  
  \centering
  \iffullversion%
    \begin{tikzpicture}[scale=\figurescale]
      \begin{axis}[
          xmin=6,
          xmax=18,
          ymode=log,
          ymin=1e-6,
          ymax=3,
          xlabel={$T$},
          no markers,
          legend pos=north east,
          legend cell align={left}
        ]
        \addplot[
          blue,
          solid
        ] table [x=t, y=epsilon, col sep=comma]{thresholding_n=1000.csv};
        \addplot[
          red,
          ultra thick,
          dashed
        ] table [x=t, y=delta, col sep=comma]{thresholding_n=1000.csv};
        \legend{$\eps$,$\del$}
      \end{axis}
    \end{tikzpicture}
  \fi
  \begin{tikzpicture}[scale=\figurescale]
    \begin{axis}[
        xmin=60,
        xmax=120,
        ymode=log,
        ymin=1e-16,
        ymax=1,
        xlabel={$T$},
        no markers,
        legend pos=north east,
        legend cell align={left}
      ]
      \addplot[
        blue,
        solid
      ] table [x=t, y=epsilon, col sep=comma]{thresholding_n=10000.csv};
      \addplot[
        red,
        ultra thick,
        dashed
      ] table [x=t, y=delta, col sep=comma]{thresholding_n=10000.csv};
      \legend{$\eps$,$\del$}
    \end{axis}
  \end{tikzpicture}
  \iffullversion%
  \begin{tikzpicture}[scale=\figurescale]
    \begin{axis}[
        xmin=550,
        xmax=700,
        ymode=log,
        ymin=1e-20,
        ymax=1,
        xlabel={$T$},
        no markers,
        legend pos=north east,
        legend cell align={left}
      ]
      \addplot[
        blue,
        solid
      ] table [x=t, y=epsilon, col sep=comma]{thresholding_n=100000.csv};
      \addplot[
        red,
        ultra thick,
        dashed
      ] table [x=t, y=delta, col sep=comma]{thresholding_n=100000.csv};
      \legend{$\eps$,$\del$}
    \end{axis}
  \end{tikzpicture}
  \fi
  \caption{
    $\eps$ and $\del$ from \autoref{thm:thresholding} as a function of the
    threshold $T$, where $|B|=0$, $p=0.5\%$, and
    \iffullversion%
      three different values of $n$: $n=\num{1000}$ (top), $n=\num{10000}$
      (middle), and $n=\num{100000}$ (bottom).
    \fi%
  }\label{fig:thresholding}
\end{figure}
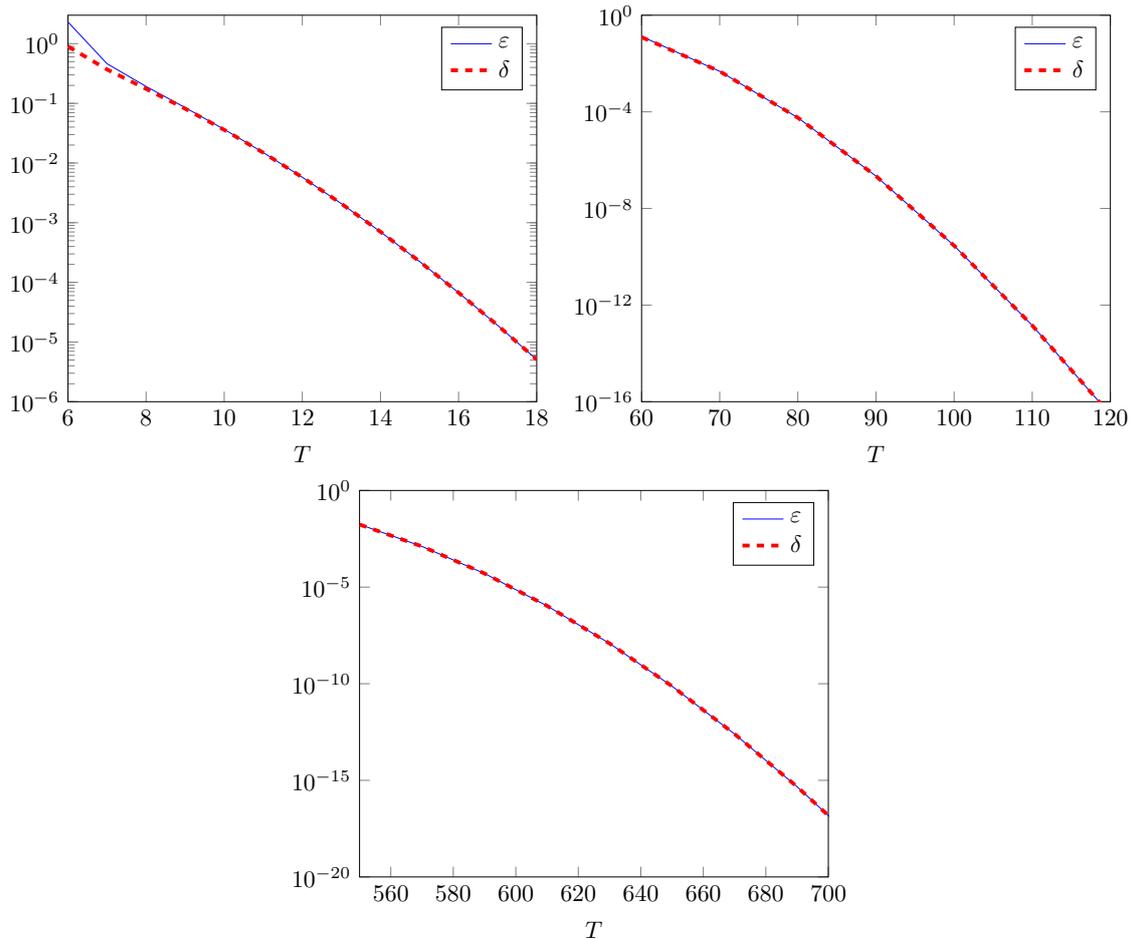

As shown in Figure~\ref{fig:thresholding}, when the threshold is above the
expected value, the values $\eps$ and $\del$ given by
Theorem~\ref{thm:thresholding} are very close. Moreover, for large $n$, these
values are extremely small. This shows that thresholding counts constitutes a
good practice, which can be used to meaningfully improve user privacy without
having to know about the data distribution in advance, like in the usage
statistics example at the beginning of this section.

A practitioner can apply the following reasoning: for each possible sequence of
actions captured by the system collecting app usage statistics, either many
users are likely to have a value of 1, in which case
Theorem~\ref{thm:binary-voting} applies and thresholding will likely not impact
data utility; or the vast majority of users will have a value of $0$, in which
case Theorem~\ref{thm:thresholding} applies and thresholding will protect the
rare users whose value is $1$.

What if the attacker has non-zero partial knowledge, but is able to interact
with the system? We saw in Example~\ref{ex:thresholding} that if this partial
knowledge is larger than the threshold, the mechanism is not private. But if
this partial knowledge is small enough, then privacy is still possible: it is
equivalent to reducing the threshold for an attacker with no partial knowledge.

\begin{prop}\label{prop:thresholding-apk-partial}
  Let $\theta$ be the same distribution as in Theorem~\ref{thm:thresholding},
  with background knowledge of size $|B|\le T$. Let $\theta'$ be the equivalent
  distribution but with $n'=n-|B|$, and no partial knowledge. Then for any $\eps$
  and $\del$, $\mecha_T$ is $(\{\theta\},\eps,\del)$-APKDP iff $\mecha_{T-|B|}$
  is $(\{\theta'\},\eps,\del)$-PPKDP\@.\iffullversion%
  
    \begin{proof}
          In this case, $\APK_{i,a,b,B}$ only depends on $T-b$ ($b$ being the number
    of ones $b$ in $B$). The same applies for $n$ and $|B|$, which only appear
    as $n-|B|$. Thus, $\mecha_T$ is $(\{\theta\},\eps,\del)$-APKDP iff
    $\mecha_{T-|B|}$ is $(\{\theta'\},\eps,\del)$-APKDP\@. As APKDP and PPKDP
    are the same when there is no background knowledge, the statement follows.

    \end{proof}
  \fi
\end{prop}
\vspace{-1em}

\subsection{Application to $k$-anonymity}\label{sec:kanon}

Sections~\ref{sec:counting} and~\ref{sec:thresholding} formalize two intuitive
phenomena under partial knowledge. First, if the attacker has a significant
enough uncertain about enough people, counting queries do not leak too much
information about individuals. Second, for counting queries that apply to rare
enough behavior, thresholding provides meaningful protection against a passive
attacker. This suggests a link to an older anonymization notion: $k$-anonymity.
In this section, we formalize that link, and combine these two intuitions to
provide a relation between $k$-anonymity and differential privacy under partial
knowledge. 

$k$-anonymity, introduced
in~\cite{samarati2001protecting,sweeney2002kanonymity}, requires each record in
a database to be indistinguishable from at least $k-1$ other records. The
intuition is that blending in a large enough crowd provides protection; this
intuition is close to the results of Section~\ref{sec:counting}. $k$-anonymity
is generally obtained by generalizing the data to group similar records
together, then dropping the groups with less than $k$ records. The link with the
results of Section~\ref{sec:thresholding} is obvious.

To formalize it, we need to clarify the notion of a $k$-anonymity mechanism. For
simplicity, we will simply assume that such a mechanism groups records by their
value, and returns a truncated histogram, where all values with less than $k$
records have been removed.

\begin{definition}[$k$-anonymity mechanism]
  The \emph{$k$-anonymity mechanism} $\mecha_k$ takes a dataset in $\datasets$
  as input, and returns a histogram in ${(\numbers\cup\bot)}^\tuples$. For each
  $t\in\tuples\cup\{\bot\}$, $\mecha_k(D)$ is defined as:
  \begin{itemize}
    \item for all $t\in\tuples$,
      $\mecha_k(D)(t)=\left|\left\{i\mid|D(i)=t\right\}\right|$ if this number
      is at least $k$ (if there are less than $k$ records with value $t$ in
      $D$);
    \item $\mecha_k(D)(t)=\bot$ otherwise.
  \end{itemize}
  If an input record is not in $\tuples$, it is ignored by $\mecha_k$.
\end{definition}

Note that we skipped the generalization step. The results below can be easily
extended to any \emph{fixed} generalization strategy, i.e.\ a fixed mapping
between $\tuples$ and an arbitrary space forming the support of the histogram.
It is important that this strategy is fixed. If this function depends on the
data, arbitrary correlations can be embedded in the output, which might leak
additional information; \emph{minimality attacks}~\cite{wong2009anonymization}
provide an example of this phenomenon.

Now, under which condition is such a mechanism private? The distribution that
captures the attacker's uncertainty must be such that for all possible values
$t\in\tuples$, either this value is rare enough to be thresholded with high
probability, either there is sufficient randomness in the input data that
releasing the exact value does not leak too much information.

In addition, we assume that it is possible for a given record to have the value
$\bot$, representing their \emph{absence} in the dataset. The count
corresponding to $\bot$ are never released. We discuss later the importance of
such a special value, and its practical interpretation.

\begin{theorem}\label{thm:kanon}
  Let $\theta$ be a distribution that generates $n$ independent records in
  $\tuples\cup\{\bot\}$. Assume that there is a $\lambda$ such that for all
  $t\in\tuples$:
  \begin{itemize}
    \item either for all indices $i$, $\proba{D(i)=t}\le \lambda$,
    \item or for all indices $i$, $\lambda\le\proba{D(i)=t}\le1-\lambda$;
  \end{itemize}
  furthermore, assume that for all indices $i$,
  $\lambda\le\proba{D(i)=\bot}\le1-\lambda$, and that the attacker does not have
  any background knowledge.

  Let $T$ be a threshold such that $r=\frac{\lambda(n-1)}{(1-\lambda)k}<1$. Then
  $\mecha_T$ is $(\{\theta\},\eps,\del)$-APKDP for all $\del\ge\del_0$, where:
  \begin{align*}
    \del_0 & = \frac{2 \cdot f(T,n-1,\lambda)}{1-r} \\
    \eps   & = 2\cdot\max\left(-\ln\left(1-\frac{f(T,n-1,\lambda)}{1-r}\right),\eps_c\right)
  \end{align*}
  and
  \iffullversion%
    $\eps_c$ is such that
  \fi%
  $\del\ge\proba{\frac{X}{Y}\ge\eps_c}$, where $X$ and
  $Y$ are two independent random variables sampled from a binomial distribution
  with $n-1$ trials and success probability $2\lambda$.
  \iffullversion%
    \begin{proof}
          For a given index $i$ and a possible record $t\in\tuples$, we compare the
    events $D(i)=t$ and $D(i)=\bot$. If we find $\eps$ and $\del$ such
    that $\mech{D}\cond{D(i)=t}\indepsdel\mech{D}\cond{D(i)=\bot}$, then we
    have $\mech{D}\cond{D(i)=t}\ind_{2\eps,2\del}\mech{D}\cond{D(i)=t}$ for
    all $t,t'\in\tuples$, which would conclude the proof immediately.

    We must consider two options: either for all indices $i$, $\proba{D(i)=t}\le
    \lambda$, or for all $i$, $\lambda\le\proba{D(i)=t}\le1-\lambda$.

    In the first case, we can reuse the proof of Theorem~\ref{thm:thresholding}:
    with probability $1-\del_0=1-\frac{f(T,n-1,\lambda)}{1-r}$, the result is
    thresholded, and the privacy loss is bounded by
    $\eps_0=-\ln\left(1-\frac{f(T,n-1,\lambda)}{1-r}\right)$. Importantly,
    comparing $D(i)=t$ and $D(i)=\bot$ allows us to restrict our analysis to the
    value of $\mecha_T(D)(t)$: for all $t'\neq t$, $\mecha_T(D)(t')$ is the same
    when $\theta$ is conditioned on $D(i)=t$ or $D(i)=\bot$.

    In the second case, we reuse the proof of Theorem~\ref{thm:binary-voting}:
    the distribution of $\mecha_T(D)(t)$ can be seen as the sum of records, each
    of whom has been randomized using a binary randomized response with
    parameter $2\lambda$. As $\mecha_T(D)(\bot)$ also follows this binary
    randomized response process, we can directly apply the proof with $\del_0$.

    Combining both cases leads to the desired result, using the
    indistinguishability between $\mecha_T(D)(t)$ and $\mecha_T(D)(\bot)$ to get
    the result between arbitrary $t$ and $t'$.

    \end{proof}
  \fi
\end{theorem}

Theorem~\ref{thm:kanon} is relatively complex, and depends on a number of
conditions. Let us discuss its limitations. Some of them are necessary for the
result to be true, others could be overcome with a more careful analysis, at the
cost of simplicity.

First, we assume that the attacker has no partial knowledge over the data. The
result can easily be extended to the case where the attacker has non-zero
\emph{passive} partial knowledge of $m$ records over the data: for the counting
case, we can simply remove these $m$ records and obtain the results with $n-m$
instead of $m$, and for the thresholding case, we can apply
Theorem~\ref{thm:thresholding} directly. The discussion in
Section~\ref{thm:thresholding}, shows that cannot be easily extended to the case
where the attacker has the ability to influence the data, unless a very small
number of records can be influenced (as in
Proposition~\ref{prop:thresholding-apk-partial}). This captures the correct
intuition that $k$-anonymity is vulnerable against active attackers.

Second, the choice distribution $\theta$ might seem artificial, carefully chosen
so the previous results can be applied. Why would there be a value $\lambda$
such that all records have a probability lower than $\lambda$ of being in a
fixed category, or larger than $\lambda$? The first option is reasonable: many
real-life distributions are long-tailed; some types of actions, or
characteristics, are simply very rare. The second option is less natural: maybe
a characteristic that is common for many people is extremely rare in others, so
requiring \emph{all} records to have a high enough probability for this record
seems too restrictive. However, note that this high probability captures the
attacker's \emph{uncertainty}: if the attacker knows that some records have a
particularly low probability of having a certain record, it is possible to
over-approximate this knowledge, and simply consider these records as known by
the attacker. We can then use the previous point to still get an upper bound on
the attacker's information gain.

Third, what is the meaning of the $\bot$ special case, and is it necessary for
the proof of Theorem~\ref{thm:kanon} to work? We use it to prove the desired
indistinguishability property in the second case of the proof. Without it, it
turns out that subtle problems can arise. Suppose, for example, that
$\tuples=\{a,b,c\}$, and that for all $i$, $\proba{D(i)=a}$ is infinitesimally
small, while $\proba{D(i)=b}$ and $\proba{D(i)=c}$ are both close to $0.5$. If
the total number of records is fixed (and implicitly assumed to be known by the
attacker), note that thresholding the count for $a$ is pointless: with high
probability, we can retrieve it by computing the difference between $n$ and the
counts for $b$ and $c$. This phenomenon is a real vulnerability of $k$-anonymity
when the total number of participants is known: any result showing that
$k$-anonymity protects privacy under partial knowledge must find a way of
guaranteeing that this does not happen.

Creating an artificial category $\bot$ whose count is never released solves this
problem, assuming that this category has sufficient uncertainty. This hides the
total number of participants and mitigate this vulnerability. Another way would
be to impose that the distribution $\theta$ has multiple $t\in\tuples$ whose
counts will likely be thresholded, and that these $t$ together have enough
uncertainty to hide the total count. This is also realistic in practice, given
that most distributions are long-tailed, but would likely require a more complex
analysis, as well as complicate the theorem statement.


Note that a link between $k$-anonymity and differential privacy was already
introduced in~\cite{li2011provably}. We use the same notion of a $k$-anonymity
mechanism, however, we model the attacker's partial knowledge differently.
In~\cite{li2011provably}, the attacker is assumed to know the value of every
single record from the original dataset, but not which records have been
randomly sampled from it. Arguably, the only way to satisfy that assumption in
practice is to have the mechanism \emph{actually} sample the data before
applying $k$-anonymity. In that case, the original differential privacy
definition is satisfied. By contrast, our setting assumes an attacker that has
some uncertainty about the value of the records themselves; we argue that this
is a much more natural way of capturing the natural assumption that the attacker
has partial knowledge over the data.


\section{Composition}

Composition theorems enable the modular analysis of complex systems and the
continued usage of mechanisms over time. In this section, we study two kinds of
composition. \emph{Sequential} composition, where multiple mechanisms are
applied to the same data, and \emph{nested} composition, where post-processing
noise is added to the result of the aggregation.

\subsection{Sequential composition}

We saw in the previous section that noiseless mechanisms could be private under
partial knowledge. For such mechanisms, composition does not hold in general. We
explain why dependencies between mechanisms are the root cause of composition
failing, and we explain how bounding this dependencies allow us to derive usable
composition results. First, we show that noiseless composition fails in general.

\begin{example}\label{ex:no-composition}
  Going back to the voting example, consider the queries ``How many people voted
  1?'' and ``How many people who are not $t$ voted 1?'', for some individual
  $t$. As shown in Section~\ref{sec:counting}, each query can be private on its
  own. However, publishing both results reveals $t$'s vote: the composition of
  both queries is not private.
\end{example}

Are there special cases where noiseless counting queries can be composed? In
this section, we propose a criterion, \emph{$(\mu,\nu)$-boundedness}, under
which sequential composition does hold.

The core problem with Example~\ref{ex:no-composition} is that the two queries
are heavily \emph{dependent} on each other. In fact, knowing the result to the
first query only leaves two options for the result of the second query: it
drastically reduces an attacker's uncertainty about the second query's result.
We show that this dependency between queries is the main obstacle towards a
composition result and prove that mechanisms where the dependency is
\emph{bounded} (\autoref{def:bounded-dependency}) can actually be
composed (\autoref{thm:sequential-composition}).


How can we formalize the \emph{bounded dependency} between mechanisms? A natural
approach is to quantify how much the additional knowledge of the first mechanism
impacts the privacy loss of the second mechanism.


\begin{definition}\label{def:dependency}
  Given two mechanisms $\mecha_1$ and $\mecha_2$, two $a,b\in \tuples$, two
  outputs $O_1,O_2$, a distribution $\theta$, an index $i$, a possible value of
  the background knowledge $\hatB$ compatible with $D(i)=a$ and $D(i)=b$, the
  \emph{dependency of $\mecha_2$ on $\mecha_1$ to distinguish $D(i)=t$ and
  $D(i)=t'$} is the function
  $\outputs^2\rightarrow\mathbb{R}\cup\set{-\infty,\infty}$ defined by:
  \begin{align*}
    & \Depab \\
    & \; = \privlossit{a}{b}{\mecha_2}\left(O_2,\left(\hatB,\mecha_1(D)=O_1\right)\right)
      - \privlossit{a}{b}{\mecha_2}\left(O_2,\hatB\right)
  \end{align*}
  using the convention $\pm\infty-x=\pm\infty$ for all $x$. 
\end{definition}

Intuitively, this value quantifies the amount of additional information that
$\mecha_1$ gives the attacker when analyzing the privacy loss of $\mecha_2$. In
Example~\ref{ex:no-composition}, the first term is $\pm\infty$, as knowing both
the results of $\mecha_1$ and $\mecha_2$ leaks the value of $D(i)$, while the
second term is typically finite. So $\Depab$ takes infinite values, which
captures the fact that the two mechanisms together leak a lot of information.

Bounding this dependency can be done in the same way as using the PLRV to define
differential privacy: we bound $\Dep$ by $\mu$ almost everywhere, and use a
small quantity $\nu$ to capture rare events where $\Dep>\mu$.

\begin{definition}[$(\mu,\nu)$-bounded dependency]\label{def:bounded-dependency}
  Given a family of distributions $\Theta$, two mechanism $\mecha_1$ and
  $\mecha_2$ are \emph{$(\mu,\nu)$-bounded dependent for $\Theta$} if for all
  $\tit$, all indices $i$ and records $a,b\in\tuples$, and all $\hbik$:
  \begin{align*}
  \expectu{\footnotesize
    \begin{matrix}\theta\cond{D(i)=a,\hatB},\\O_1\sim\mecha_1(D),\\O_2\sim\mecha_2(D)\end{matrix}
  }{\max\left(0,1-e^{\mu-\Depab}\right)}
  \end{align*}
  is smaller or equal to $\nu$.
\end{definition}

This notion formalizes the intuition that the result of the first mechanism
should not impact ``too much'' the result of the second mechanism. As we show in
the following theorem, the dependency of $\mecha_2$ on $\mecha_1$ can be used to
express the PLRV of the composed mechanism as a function of the PLRV of the two
original mechanisms. As a direct consequence, we show that two
$(\mu,\nu)$-bounded dependent mechanisms can be sequentially composed.

\begin{theorem}\label{thm:sequential-composition}
  Given a distribution $\theta$, two mechanisms $\mecha_1, \mecha_2$, an indice
  $i$, records $a,b\in\tuples$, and $\hbik$, the PLRV of the composed mechanism
  $\mecha(D):=(\mecha_1(D),\mecha_2(D))$ satisfies:
  \iffullversion%
  \begin{align*}
    & \privlossMit{a}{b}\left(O,\hatB\right)\\
    & \; = 2\cdot\Depab \\
    & \qquad + \privlossit{a}{b}{\mecha_1}\left(O_1,\hatB\right) + \privlossit{a}{b}{\mecha_2}\left(O_2,\hatB\right).
  \end{align*}
  \fi
  As a corollary, if $\mecha_1, \mecha_2$ are $(\mu,\nu)$-bounded dependent,
  if $\mecha_1$ is $(\Theta,\eps_1,\delta_1)$-APK, and if $\mecha_2$ is
  $(\Theta,\eps_2,\delta_2)$-APK for $\Theta$, then $\mecha$ is
  $(\Theta,2\mu+\eps_1+\eps_2,\del_1+\del_2+\nu)$-APK\@.

  \iffullversion%
    \begin{proof}
    To prove the main statement, we decompose:
    \begin{align*}
      & \privlossMit{a}{b}(O,\hatB)\\
      & \; = \ln\frac{%
          \probasc{\theta}{\mecha(D) = (O_1,O_2)}{D(i)=a, \hatB}
        }{%
          \probasc{\theta}{\mecha(D) = (O_1,O_2)}{D(i)=b, \hatB}
        } \\
      & \; = \ln\frac{%
      \probasc{\theta}{\mecha_2(D) = O_2}{\mecha_1(D) = O_1, D(i)=a , \hatB}}{%
      \probasc{\theta}{\mecha_2(D) = O_2}{\mecha_1(D) = O_1, D(i)=b, \hatB}}\\
      & \; \; \; \; + \ln\frac{%
      \probasc{\theta}{\mecha_1(D) = O_1}{D(i)=a, \hatB}}{%
      \probasc{\theta}{\mecha_1(D) = O_1}{D(i)=b, \hatB}}
    \end{align*}

    and use the definition of $\Depab$.


    To prove the composition theorem, we have to show that, if we denote
    $\eps=\eps_1+\eps_2+\mu$ and $\del=\del_1+\del_2+\nu$:
    \begin{align*}
      \expectu{\theta\cond{D(i)=a,\hatB},O\sim\mech{D}}{\max\left(0,1-e^{\eps-\privlossMit{a}{b}(O,\hatB)}\right)}\le\del.
    \end{align*}
    The function $f(x)=\max\left(0,1-e^x\right)$ satisfies ``for all $x$
    and $y$, $f(x+y)\le f(x)+f(y)$'': if $x>0$, then $f(x)=0$, and
    $1-e^{x+y}\le1-e^y$. The same holds when $y>0$. If $x\le0$ and $y\le0$, then
    we must show that $1-e^{x+y}\le2-e^x-e^y$, which is equivalent to
    $\left(e^x-1\right)\left(e^y-1\right)\ge0$.

    We then define $a=\eps_1-\privlossit{a}{b}{\mecha_1}\left(O_1,\hatB\right)$,
    $b=\eps_2-\privlossit{a}{b}{\mecha_2}\left(O_2,\hatB\right)$, and
    $c=\mu-\Depab$, and we use subadditivity: $f(a+b+c)\le f(a)+f(b)+f(c)$. As
    $a+b+c=\eps-\privlossit{a}{b}{\mecha}\left(O,\hatB\right)$, we can directly
    plug this into the expression above and use the theorem assumption to prove
    the $\del_1+\del_2+\nu=\del$ bound.

    \end{proof}
  \fi
\end{theorem}

Note that the characterization of $\privlossMit{a}{b}\left(O,\hatB\right)$
enables the use of more sophisticated composition bounds for differential
privacy, such as the advanced composition theorem~\cite{dwork2014algorithmic},
R{\'e}nyi Differential Privacy~\cite{mironov2017renyi}, or privacy
buckets~\cite{meiser2018tight}. For simplicity, the proof uses the standard
(non-tight) composition bound for DP\@.

A common special case directly leads to $(0,0)$-bounded dependent mechanisms:
two mechanisms that work on distinct parts of a database are $(0,0)$-bounded
dependent if these two parts are independent.

\begin{prop}\label{prop:independent-mechanisms}
  Let $\Theta$ be a family of distributions, and let $\mecha_1$ and $\mecha_2$
  be mechanisms. Assume that for any $\tit$, there are functions $\pi_1$ and
  $\pi_2$ such that $\pi_1(D)\cond{D\sim\theta}$ and
  $\pi_2(D)\cond{D\sim\theta}$ are independent, and functions $\mecha^\prime_1$
  and $\mecha^\prime_2$ such that
  $\mecha_1(D)=\mecha^\prime_1\left(\pi_1(D)\right)$ and
  $\mecha_2(D)=\mecha^\prime_2\left(\pi_2(D)\right)$. Then $\mecha_1$ and
  $\mecha_2$ are $(0,0)$-bounded dependent.
\end{prop}

We now present an natural example of a practical scenario where we can use these
composition results.

\begin{example}\label{ex:composition}
  Consider a regularly updated database, like usage information about an online
  service. Statistics $q$ are computed from this database: for example, among
  registered users, how many of them used a specific feature on any given day.
  This count is released daily, and we want to understand how the privacy of a
  particular user is impacted over time.

  This can be represented by a database $D$ where each record $i$ is a series of
  binary values ${\left(D(i)\right)}_j$, where $j=0,1,2,\cdots$, and we release
  a series of mechanisms $\mecha_j(D)=\sum_i q({D(i)}_j)$. The results of
  Section~\ref{sec:applications} can be used to determine the privacy of each
  $\mecha_j$ depending on the data-generating distribution $\theta$. The goal is
  to determine the privacy of multiple queries, assuming independence between
  $D\left(i_1\right)$ and $D\left(i_2\right)$ for all $i_1\neq i_2$.

  The analysis of the privacy guarantees offered by this setting over time
  depends on $\theta$, and on the correlations between the different values of a
  record. If ${D(i)}_{j_1}$ is independent from ${D(i)}_{j_2}$ for all $j_1\neq
  j_2$, then the result is direct. Otherwise, we must quantify the maximum
  amount of correlation between ${D(i)}_j$ and ${D(i)}_{j+1}$. Quantifying this
  can be done using indistinguishability: we can assume, for example, that there
  is a $c\ge0$ such that for all $a\in\tuples$ and all indices $i$ and $j$:
  \begin{align*}
    \left({D(i)}_{j+1}\right)\cond{{D(i)}_j=a} \ind_c {D(i)}_j.
  \end{align*}
  Under this assumption, it is easy to verify that mechanisms $\mecha_j$ and
  $\mecha_{j+1}$ are $(2c,0)$-bounded dependent, so we can use the composition
  result of Theorem~\ref{thm:sequential-composition} and derive bounds on the
  privacy leakage over time.
\end{example}

This approach can be extended to other scenarios, for example if only a subset
of users participate to each update, or if a referendum contains multiple
questions, whose answers are correlated. Another possible scenario is if only a
subset of users participate to each update. We can represent this by having
${D(i)}_j$ be either a categorical value (which encodes e.g.\ the type of
interaction) or a special value $\bot$ that encodes ``user $i$ did not
participate to this update''. The probabilities and correlation relationships of
different values associated with the same user can be set to capture different
scenarios (e.g.\ the probability that ${D(i)}_j=\bot$ can be large, to capture a
scenario where few users participate every round).



\subsection{Composing partial knowledge with post-processing noise}

The results of Sections~\ref{sec:counting} to~\ref{sec:kanon} show that
noiseless mechanisms can be considered private, assuming some additional
assumptions on the attacker's background knowledge. With enough records, even
pessimistic assumptions (considering an attacker who knows a large fraction of
records) can lead to very small values of $\eps$ and $\del$. However, one could
still consider these assumptions as too brittle, and decide to add a small
amount of additional noise to the mechanism to have it satisfy differential
privacy in its original form.

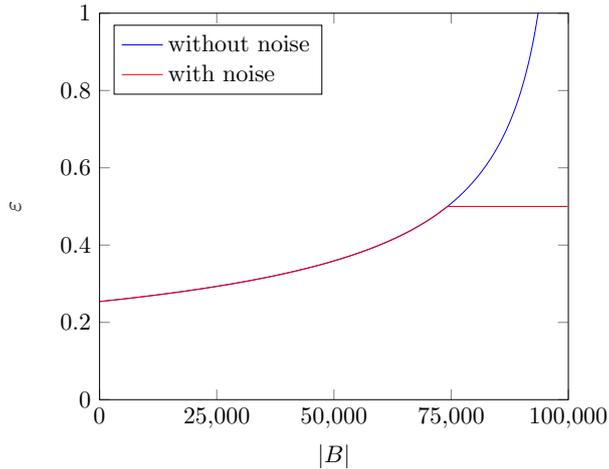
\begin{figure} 
  \centering
  \begin{tikzpicture}[scale=\figurescale]
    \begin{axis}[
        /pgf/number format/fixed,
        scaled ticks=false,
        tick label style={/pgf/number format/fixed},
        xmin=0,
        xmax=100000,
        xtick={0,25000,50000,75000,100000},
        ymin=0,
        ymax=1,
        xlabel=$|B|$,
        ylabel=$\eps$,
        legend pos=north west,
        legend cell align={left}
      ]
      \addplot[
        blue,
        solid,
      ] table [x=b, y=epswithout, col sep=comma]{noisefloor.csv};
      \addplot[
        red,
        solid
      ] table [x=b, y=epswith, col sep=comma]{noisefloor.csv};
      \legend{without noise,with noise}
    \end{axis}
  \end{tikzpicture}
  \caption{$\eps$ from the closed-form formula of \autoref{thm:binary-voting}
    for $\del=10^{-10}$, $\lambda=0.05$, and $n=\num{100000}$, as a function of
    the number of records known by the attacker $|B|$. We compare two scenarios:
    either we do not add any post-processing noise, or we add Laplace noise of
    scale 2 to the output.}\label{fig:noisefloor}
\end{figure}

Such mechanisms have a \emph{double} privacy guarantee: under realistic
assumptions, their privacy level is very high thanks to the attacker's
uncertainty, and the additional noise provides a ``worst-case'' privacy level
that the mechanism satisfies independently of the attacker capabilities. Without
noise, we can use results like Theorem~\ref{thm:binary-voting} to show that a
given aggregation over $n$ records is
$\left(\Theta(|B|),\eps(|B|),\del(|B|)\right)$-APKDP (or PPKDP), where $|B|$
is the number of records that the attacker knows. In situations like ones we
have seen so far, $\eps(|B|)$ and $\del(|B|)$ can be very small when $|B|$ is
close to $0$, but might become unacceptably high when $|B|$ gets close to $n$.
Adding noise can be a way to guarantee that $\eps(|B|)$ and $\del(|B|)$ never
get above a certain point: when there is not enough randomness coming from the
data anymore, the guarantee from post-processing noise take over.
Figure~\ref{fig:noisefloor} illustrates this phenomenon.

It is also natural to wonder whether the two sources of uncertainty
could be combined. The privacy guarantees from Theorem~\ref{thm:binary-voting}
come from the shape of the binomial distribution, just like the shape of Laplace
noise is the reason why adding it to the result of an aggregation can provide
$\eps$-DP\@. It seems intuitive that combining two sources of noise would have a
\emph{larger} effect.

\begin{figure}
  \centering
  \begin{tikzpicture}[scale=\figurescale]
    \begin{axis}[
        xmin=0.2,
        xmax=1,
        ymin=1e-14,
        ymax=1e-2,
        ymode=log,
        ytick={1e-14, 1e-12, 1e-10, 1e-8, 1e-6, 1e-4, 1e-2},
        xlabel=$\eps$,
        ylabel=$\del$,
        legend pos=north east,
      ]
      \addplot[
        blue,
        dashed
      ] table [x=epsilon, y=delta, col sep=comma]{nestedcomposition-1000-0.05-0.csv};
      \addplot[
        orange,
        solid
      ] table [x=epsilon, y=delta, col sep=comma]{nestedcomposition-1000-0.05-0.5.csv};
      \addplot[
        red,
        solid
      ] table [x=epsilon, y=delta, col sep=comma]{nestedcomposition-1000-0.05-0.6.csv};
      \legend{no noise,$p=0.5$,$p=0.6$}
    \end{axis}
  \end{tikzpicture}
  \caption{Numerical computation of the $\epsdel$ bounds given by
    \autoref{thm:binary-voting} with $n=\num{10000}$ and $\lambda=0.05$ (in
    dashed blue), compared to the bounds obtained by adding two-sided
    geometric noise of parameter $p=0.5$ or $p=0.75$
    \iffullversion%
      (see \autoref{def:geometric})
    \fi
    and combining both distributions.}%
  \label{fig:nested-composition}
\end{figure}
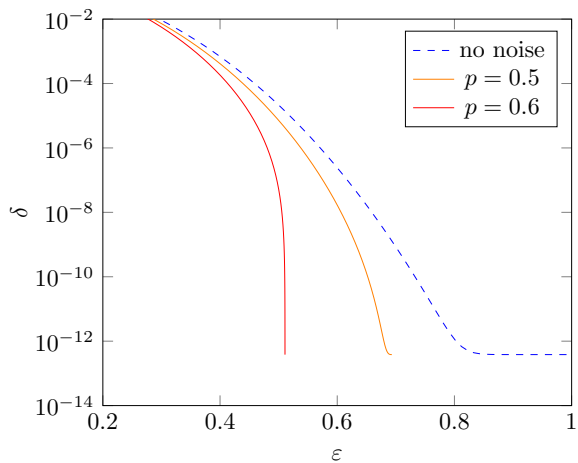

In some cases, this effect can be numerically estimated. Given a noise
distribution $X$ added to a mechanism of sensitivity $s$, the PLRV can be
obtained by comparing the distributions of $X$ and $X+s$. To estimate the PLRV
coming from two noise sources summed together (for example, binomial and
geometric noise), we can simply compute the convolution of the corresponding two
distributions, and use the result to compute the PLRV, and thus, the $\epsdel$
graph. We demonstrate this approach in \autoref{fig:nested-composition}, where
we add two-sided geometric noise to a noiseless counting query.

\iffullversion%
  \begin{definition}[Two-sided geometric distribution]\label{def:geometric}
    The \emph{two-sided geometric distribution of mean 0 and of parameter
    $p\in(0,1)$} is the probability distribution such that a random variable $X$
    sampled from the distribution follows, for all integers $k$:
    \[
      \proba{X=k} = \frac{1-p}{1+p}p^{|k|}.
    \]
  \end{definition}
\fi

It is natural to ask whether we could obtain generic results that quantify the
combined effect of noise coming from the input data and noise added after the
aggregation mechanism, without numerical evaluation.
In~\cite{grining2017towards}, the authors propose such a result, based on the
fact that Gaussian distributions are closed under convolution. The noise from
the input data is approximated by a Gaussian using the central limit theorem,
and their Theorem~6 shows that adding Gaussian noise leads to a smaller $\eps$.
However, since the $\del$ term comes from the central limit theorem
approximation, it cannot be improved beyond $\del=O(1/\sqrt{n})$ in general.

We could solve this by simply making the assumption that the input data unknown
from the attacker \emph{actually} follows a Gaussian distribution. Sadly, the
corresponding result would be very brittle: if an attacker does not conform
exactly to this approximation, then the result no longer holds. This is a major
criticism of privacy definitions which assume the input data has inherent
randomness~\cite{steinke2020pitfalls}. The results of this paper are not so
brittle, as the privacy guarantees degrade gracefully with the assumptions we
make on the attacker's partial knowledge (e.g.\ the number of records known, or
the value of $p$ in Theorems~\ref{thm:binary-voting} or~\ref{thm:thresholding}).

Another approach would be to choose the noise added as post-processing based on the
natural noise distributions emerging from the partial knowledge assumption. For
example, since the proof of Theorem~\ref{thm:binary-voting} uses the fact that
the attacker uncertainty corresponds to binomial noise, we could also add
binomial noise as post-processing, since $B(n,p)+B(m,p)=B(n+m,p)$. Yet, this
property depends on the exact value of $p$, which creates a brittleness we
were trying to avoid.

The question of computing the privacy loss in situations where multiple sources
of randomness are combined appears in other scenarios. Amplification by sampling
or amplification by shuffling are examples of such results. These two classes of
results are \emph{generic}: they do not depend on the exact mechanism used to
obtain the initial $\epsdel$-DP guarantee. It is unlikely that such generic
results exist when combining two arbitrary sources of noise, each of which
satisfies $\epsdel$-DP\@.

Other results depend on additional assumptions on the noise distribution,
like amplification by iteration~\cite{feldman2018privacy} or amplification by
mixing and diffusion mechanisms~\cite{balle2019amplification}. These do not seem
to bring significant improvements in scenarios like \autoref{thm:binary-voting}
with post-processing noise: amplification by iteration characterizes adding noise
many times (not only once), while amplification by mixing
and diffusion requires stronger assumptions on the original noise distribution.

An generic result on the privacy guarantee of chained $\eps$-DP mechanisms
appears in~\cite{erlingsson2020encode} (Appendix~B). This tight result is only
valid for pure $\eps$-DP, but the main building block holds for $\epsdel$-DP
mechanisms: proving a fully generic chained composition result is equivalent to
solving the special case where the input and output of both mechanisms have
values in $\set{0,1}$. This result can likely be extended to the $\epsdel$-DP,
although the analysis is surprisingly non-trivial, and fully generic optimality
results do not necessarily mean optimality for the special case of additive
noise mechanisms.

\section{Conclusion}

\iffullversion%
We identified issues that arise with existing definitions in the presence of
correlations in the data. We proposed a criterion that resolves these issues and
unifies different approaches, and we showed that an attacker with partial
knowledge can be either passive or active.
\fi%
We delineated these cases in two definitions, and we proved fundamental results
about these definitions. We then quantified the privacy guarantees of natural,
practical mechanisms, under realistic assumptions. We improved known results on
the noiseless privacy of counting queries, and showed that thresholding can
protect the privacy of individuals even in cases where is little randomness in
the original data. Finally, we showed a natural relationship linking how
correlated two mechanisms are to how well their privacy guarantees compose, and
proposed initial results on nested composition. We hope that this work will
encourage the privacy analysis of different natural mechanisms under partial
knowledge, extending this kind of analysis to systems more complex than
individual queries.


\bibliographystyle{alpha}
\bibliography{biblio}

\iffullversion%
\fi


\end{document}